\documentclass[%
    reprint,
    superscriptaddress,
    amsmath,amssymb,
    aip,
    jcp,
    floatfix,
    letterpaper,
]{revtex4-2}

\usepackage{graphicx}% Include figure files
\usepackage[export]{adjustbox}
\usepackage[caption=false]{subfig}
\usepackage{dcolumn}% Align table columns on decimal point
\usepackage{bm}% bold math
\usepackage{braket}
\usepackage{comment}
\usepackage{xr} % cross reference

\usepackage[english]{babel}
\usepackage[T1]{fontenc}
\usepackage[version=4]{mhchem}
\usepackage{balance}
\usepackage[caption=false]{subfig}
\usepackage{placeins}
\usepackage{multirow}
\usepackage{xcolor}

\PassOptionsToPackage{hyphens}{url}
\usepackage[hyperindex,breaklinks]{hyperref}  
\newcommand{\red}[1]{#1}

\begin{document}
\preprint{APS/123-QED}

\title{Deriving effective electrode-ion interactions from free-energy profiles at electrochemical interfaces}

\author{Fabrice Roncoroni}
    %\email{roncoroni@lbl.gov}
    \affiliation{The Molecular Foundry, Lawrence Berkeley National Laboratory, Berkeley, California 94720, United States}
\author{Abrar Faiyad}
    %\email{afaiyad@ucmerced.edu}
    \affiliation{University of California Merced, Merced, California 95343, United States}
\author{Yichen Li}
    %\email{yli330@ucmerced.edu}
    \affiliation{University of California Merced, Merced, California 95343, United States}
\author{Tao Ye}
    %\email{tye2@ucmerced.edu}
    \affiliation{University of California Merced, Merced, California 95343, United States}
\author{Ashlie Martini}
    %\email{amartini@ucmerced.edu}
    \affiliation{University of California Merced, Merced, California 95343, United States}
\author{David Prendergast}
    \email{dgprendergast@lbl.gov}
    \affiliation{The Molecular Foundry, Lawrence Berkeley National Laboratory, Berkeley, California 94720, United States}
%\collaboration{Funding for Accelerated, Inclusive Research (FAIR)}%\noaffiliation

\date{\today}% It is always \today, today,
             %  but any date may be explicitly specified

\begin{abstract}
Understanding ion adsorption at electrified metal-electrolyte interfaces is essential for accurate modeling of electrochemical systems. Here, we systematically investigate the free energy profiles of \ce{Na+}, \ce{Cl-}, and \ce{F-} ions at the \ce{Au(111)}-water interface using enhanced sampling molecular dynamics with both classical force fields and machine-learned interatomic potentials (MLIPs). Our classical metadynamics results reveal a strong dependence of predicted ion adsorption on the Lennard-Jones parameters, highlighting that -- without due care -- standard mixing rules can lead to qualitatively incorrect descriptions of ion-metal interactions. We present a systematic methodology for tuning the cross-term LJ parameters to control adsorption energetics in agreement with more accurate models. As a surrogate for an \emph{ab initio} model, we employed the recently released Universal Models for Atoms (UMA) MLIP, which validates classical trends and displays strong specific adsorption for chloride, weak adsorption for fluoride, and no specific adsorption for sodium, in agreement with experimental and theoretical expectations. By integrating molecular-level adsorption free energies into continuum models of the electric double layer, we show that specific ion adsorption substantially alters the interfacial ion population, the potential of zero charge, and the differential capacitance of the system. Our results underscore the critical importance of force field parameterization and advanced interatomic potentials for the predictive modeling of ion-specific effects at electrified interfaces and provide a robust framework for bridging molecular simulations and continuum electrochemical models.
\end{abstract}

%\keywords{Suggested keywords}%Use showkeys class option if keyword
                              %display desired
\maketitle

\section{\label{sec:outline}Introduction}

The electrochemical interface, the region where an electrode and an electrolyte meet, shows properties that can differ greatly from those present in a bulk electrolyte. Notably, it plays a fundamental role in regulating interfacial charge processes and ultimately chemical reactions at the interface~\cite{Schnur2009, Schwarz2020}. Describing the interfacial region with simulations is a challenging task that requires balancing the accuracy and efficiency of the calculations~\cite{Abidi2021}. Furthermore, even idealized interfaces (e.g., homogeneous fluids in contact with extended crystalline facets of solids) are inherently open systems in contact with semi-infinite regions that may be defined by thermodynamic and/or electrostatic boundary conditions, such as chemical potentials in the bulk electrolyte or applied (electronic) potentials at the electrode surface~\cite{Steinmann2021}. Modeling interfaces at the atomistic level requires using finite (countable) systems, and equilibration to states that match the boundary conditions can be challenging. Such systems can be studied with \textit{ab initio} molecular dynamics methods (AIMD) with the advantage of being able to capture polarization, charge transfer effects, bond breaking, and bond formation while requiring little parametrization. However, this higher degree of accuracy comes at the cost of limited system sizes and increased computational time. Considering that events happening at the interface occur at timescales ranging up to several nanoseconds (for example, ion diffusion) or even longer (depending on the viscosity of the fluid, adsorption at the interface, etc.), while requiring atomic-scale resolution of the order of femtoseconds, a pure \textit{ab initio} description of the electrochemical interface can easily become too expensive from a time perspective.

Classical molecular dynamics (MD) offers a computationally efficient alternative to \textit{ab initio} methods, enabling the simulation of larger systems over longer timescales~\cite{Tuckerman2000,Buehler2007,Oliveira2009,Hollingsworth2018}. This computational efficiency enables the sampling of much longer trajectories and calculation of statistically significant estimates of observables. However, this computational advantage comes at the cost of accuracy and is limited by the availability of interaction parameters. Indeed, simple all-atom classical force fields may be fundamentally incapable of describing chemical reactions, metallic behavior, and charge transfer events at the electrode-electrolyte interface. Despite their shortcomings, the ability to sample longer, larger trajectories allows classical force fields to still play a pivotal role in the investigation of electrochemical interfaces.

Recent advances in Machine-Learned Interatomic Potentials (MLIPs) trained on millions of DFT calculations on elements spanning much of the periodic table, is enabling molecular‑dynamics simulations with near‑DFT accuracy with computational speeds on par with classical MD.~\cite{Dral2020, Ko2021, Zubatiuk2021, Kulichenko2021, Kim2024-ui, Bochkarev2024-me, Yin2025-xi, Zhang2025-gh, Yang2024-kb, Xie2017-pr,Deng2023-vb}. The MLIPs help overcome many of the accuracy limitations of classical MD and may prove useful in the study of chemical reactions, electrochemical interactions, metallic behavior, etc. This higher accuracy at lower computational cost is particularly relevant for enhanced sampling techniques, where long trajectories are required to perform appropriate thermodynamic sampling to obtain a converged result. The ability to perform higher-accuracy MD simulations at a fraction of the computational cost of DFT methods is extremely promising for simulations of electrochemical interfaces.

\subsection{Computational Modeling for Electrolytes and Metal Surfaces}

There are multiple well-established and optimized force field parameters for ions in aqueous electrolytes. Well-known examples for the halide, ammonium, alkali and alkaline-earth metal ions include the works of \AA{}qvist~\cite{Aqvist1990}, Smith and Dang ~\cite{Smith1994}, Jensen and Jorgensen~\cite{Jensen2006} or Cheatham and Joung~\cite{Joung2008}. But efforts aimed at refining and optimizing the quality of the parameters are far from over~\cite{Loche2021}, and newer force field models are still being developed to improve their treatment of polarizability, application to more complicated systems, or extensions to new solvent models~\cite{Sengupta2021}.

Ionic parameters are commonly derived by either fitting measured (experimental) or calculated (\emph{ab initio}) properties so that they can be reproduced in the classical simulation. When considering ionic species in solution, this involves matching quantities such as the hydration free energy, the coordination number of the first solvation shell, the ion-water radial distribution function, the binding energy, or the ion-oxygen distance (IOD)~\cite{Li2013,Li2015,Li2015-hcmi}. The result is a set of optimized parameters and charges, which are then embedded into a classical model composed of non-polarizable electrostatic interactions and Lennard-Jones (LJ) interactions, such as the 12-6 functional form:
\begin{equation}
    \label{eq:LJ_potential}
      V_{LJ,ij}(r) = 4\varepsilon_{ij}  \left[\left(\frac{\sigma_{ij}}{r_{ij}}\right)^{12}
        - \left(\frac{\sigma_{ij}}{r_{ij}}\right)^{6} \right]
      \enspace.
\end{equation}
where $r_{ij}$ is the separation between particle $i$ and particle $j$, $\varepsilon_{ij}$ is the depth of the Lennard-Jones potential well, and $\sigma_{ij}$ is the interparticle distance at which the Lennard-Jones potential vanishes.

To determine the interaction parameters $\varepsilon_{ij}$ and $\sigma_{ij}$ for pairs of different atom types, the single-particle parameters are often combined using mixing rules~\cite{Giri2017,Mao2023,Oliveira2023-px}. The two most commonly used rules are the \textit{geometric} (sometimes referred to as Good-Hope~\cite{Good1970}) mixing rule, which uses the geometric mean for both $\varepsilon_{ij}$ and $\sigma_{ij}$:
\begin{align}
    \varepsilon_{ij} &= \sqrt{\varepsilon_{ii} \varepsilon_{jj}} \notag \\
    \sigma_{ij} &= \sqrt{\sigma_{ii} \sigma_{jj}}  \enspace,\label{eq:geometric_mr}
\end{align}
and the \textit{arithmetic} (referred to as Lorentz-Berthelot) mixing rule, which uses the geometric mean for $\varepsilon_{ij}$ and the arithmetic mean for $\sigma_{ij}$:
\begin{align}
    \varepsilon_{ij} &= \sqrt{\varepsilon_{ii} \varepsilon_{jj}} \notag \\
    \sigma_{ij} &= \frac{1}{2} \left( \sigma_{ii} + \sigma_{jj} \right) \enspace.\label{eq:arithmetic_mr}
\end{align}
There is no universal consensus on which mixing rule is most appropriate to calculate particle interactions, and both rules appear interchangeably in the literature as authors normally adhere to the convention of the MD code or force field employed. For example, the OPLS-AA force field uses geometric mixing rules~\cite{Jorgensen1996}, while the AMBER force fields use Lorentz-Berthelot mixing rules~\cite{Chen2007}. In some cases, parameters that were originally derived with different mixing rules end up being used together within the same study, and -- without due testing -- this can affect simulation outcomes and lead to unphysical results~\cite{Chen2007}.

Alternatively, cross-atom interactions can be initialized explicitly without relying on the mixing rules. This, however, requires proper reparametrization of the interaction against some benchmark data. There are examples where efforts to explicitly reparametrize cross-atom interactions for the study of interfacial systems were conducted. Berg \textit{et al.}~\cite{Berg2017} developed a modified force field for \ce{H2O} by optimizing the \ce{Au-O} and \ce{Au-H} parameters against DFT simulations. They found that the softer Morse and Buckingham potentials -- instead of the more commonly used LJ potential -- were better suited to describe the water-\ce{Au(111)} interface. Fyta \textit{et al.}~\cite{Fyta2012} explicitly reparametrized the cation-anion interaction parameters by rescaling the standard mixing rules while preserving the ion-water interactions. Efforts towards the explicit reparametrization of the interaction between solvated ions and metallic surfaces are less frequently reported in the literature~\cite{Striolo2016}. Of particular value is the letter by Williams and coworkers~\cite{Williams2017}. In their work on the graphene-electrolyte interface, the authors showed that the attraction between the ions and the surface was underestimated unless the ion-$\pi$ interaction was included in the intermolecular potential. By tuning the 12-6 LJ ion-carbon interaction parameter $\varepsilon_{i\text{-}c}$, they were able to account for this effect and effectively model graphene-electrolyte systems within a classical framework while avoiding the need for more computationally expensive methods such as polarizable models or AIMD. A similar optimization procedure was taken by subsequent literature~\cite{Liao2023}, and studies on ion diffusion in graphene nanochannels confirm how the choice of explicit ion-graphene interaction can have a large effect on the ionic density profile~\cite{Liao2025}. With this in mind, it is clear that there is no guarantee of the transferability of the standard mixing rules to more complex systems. Specifically, parameters that were derived to describe properties of ions in bulk water might fail to describe interfaces~\cite{Sundararaman2022}, and proper reparametrization of the ion-surface interaction might be necessary.

When it comes to the description of the metallic electrode, LJ parameters for several face-centered cubic (FCC) metals, such as those developed by Heinz \textit{et al.}~\cite{Heinz2008,Kanhaiya2021}, are available in the literature. The parameters were specifically optimized to reproduce experimental densities, surface tensions, and interfacial properties with water and (bio)organic molecules, as well as mechanical properties. However, simple LJ models for metals do not account for electronic structure effects or covalent interactions. Indeed, classical models cannot explicitly reproduce the electronic response of free electrons in metals to charged species (solvent and electrolyte) and to the presence of an external field. Simulations suggest that adding polarization to electrolyte solution models has limited influence on the solvent structure of the interface~\cite{Nair2025}, yet it has been found to affect ion adsorption and wetting phenomena~\cite{Elliott2022}. While a pure \textit{ab initio} description of the electrode-electrolyte interface still challenges the limits of current computational capabilities~\cite{Scalfi2020}, different models have been developed to include the metallic character of the electrode in classical MD~\cite{Serva2021,Scalfi2021}. Geada \textit{et al.}~\cite{Geada2018} introduced a re-parametrization of the LJ terms together with an atom-centered polarizable model, treating the gold atom as a core-shell system composed of a positively charged core atom bound by a spring to a negatively charged dummy atom. This model mimics an image charge potential -- albeit based on a physical picture that more closely resembles dielectric rather than metallic screening -- and increases the tendency of charged species -- in the gas phase and to a lesser extent solvated -- to adsorb to the metal. An alternative model, first introduced by Siepmann and Sprik~\cite{Siepmann1995}, allows the charges on the metal atoms to dynamically adjust to respond to the surrounding fields and charged species while keeping the electrodes at a finite potential difference (constant potential method, CPM), which mimics the physical process of an ideal polarizable electrode~\cite{Ahrens-Iwers2022,Andersson2023}. A further implementation of the method that applies a finite field across the simulation box boundaries~\cite{Dufils2021,Tee2022} allows for fully periodic simulations at a reduced computational cost.

\red{Ionic polarization can also play an important role when it comes to ion-metal interactions, and can lead to an increased specific adsorption. The effect has been shown to be particularly pronounced for larger ions -- such as \ce{K+} or \ce{Cs+} cations -- at the graphene interface~\cite{Zhan2019}. Similarly, studies reliying on polarizable force fields for the ions to describe the water-air interface suggest that non-polarizable ions are repelled by the interface while polarizable ones tend to be attracted to it~\cite{Jungwirth2002,Jungwirth2006}. Indeed, the presence of an anisotropic interfacial environment stabilizes polarization interactions~\cite{Vrbka2004}.
}

In recent years, there has been an exponential development of MLIPs. MLIPs are capable of simulating materials with accuracy comparable to DFT methods at a fraction of the computational cost~\cite{Jacobs2025}. Moreover, some advanced MLIPs are capable of simulating a wide range of elements in different phases, overcoming the need for specialized interatomic potentials designed for specific materials. Recent advances include the development of neural network potentials and graph-based models, such as SchNet~\cite{Schutt2018}, DeepMD~\cite{Wang2018}, and NequIP~\cite{Batzner2022}. MLIPs are trained on quantum mechanical data and can capture complex many-body interactions, polarization, and even some aspects of charge transfer that are essential for modeling electrified interfaces. Efforts have also been made to explicitly train MLIPs on interfacial systems, such as metal-water and metal-electrolyte interfaces, enabling accurate simulations of water structuring, ion adsorption, and electric double layer formation at metallic surfaces~\cite{Natarajan2016,Mikkelsen2021,Guo2025}. Despite these advances, challenges remain: most conventional MLIPs are inherently short-ranged, with finite cutoffs that limit their ability to capture long-range electrostatics and the dielectric response of the solvent, all crucial features for realistic modeling of electrochemical interfaces~\cite{Zhu2025}.

Recent advances in MLIPs have led to the development of transferable models trained on extensive and diverse chemical datasets. In particular, Meta's Universal Models for Atoms (UMA)~\cite{UMA-paper}, trained on the Open Materials 2024 (OMat24)~\cite{OMat24}, Open Molecular Crystals 2025 (OMC25) and Open Molecules 2025 (OMol25) datasets~\cite{Levine2025}, represents a new generation of MLIPs designed for broad chemical coverage and high accuracy across a wide range of elements, molecular species, and condensed-phase systems. The UMA training dataset encompasses millions of quantum-mechanical reference calculations for organic molecules, ions, water, and various solid-state environments, ensuring that the model is exposed to a wide variety of local chemical environments during training. However, to fully realize the potential of these universal machine learning models, it is essential that the community systematically benchmark and validate their performance across diverse systems. Within this context, the combination of MLIPs with enhanced sampling methods allows for the exploration of rare events -- possibly absent from the training sets -- that challenge the extrapolation power of such models. At the same time, the greater performance of MLIPs allows for the extensive sampling of free energy landscapes that would have been computationally prohibitive from a purely \textit{ab initio} approach. 

\subsection{Scope of This Work}

Since ionic force field parameters are typically developed to describe bulk solvation of salts, their ability to accurately describe interfacial systems, such as the electrode-electrolyte interface, is not well-defined. In this work, we systematically investigate how the choice of ion force field parameters affects the adsorption free energy profiles of \ce{Na+}, \ce{Cl-}, and \ce{F-} at the water-Au(111) interface. Using enhanced sampling techniques~\cite{Laio2002}, we first quantify the sensitivity of interfacial ion adsorption to four different widely used classical LJ parameter sets: Jensen and Jorgensen~\cite{Jensen2006}, Joung and Cheatham~\cite{Joung2008}, Li and Merz~\cite{Sengupta2021}, and Smith and Dang~\cite{Dang1992,Smith1994}. We then discuss the effect of the 12-6 LJ parameters on the adsorption behavior and propose a simple scheme to predict changes in parameters.

Beyond classical force fields, we assess the capability of state-of-the-art MLIPs, specifically the Universal Models for Atoms (UMA)~\cite{UMA-paper, Levine2025}, to describe ion adsorption at the gold/water interface. We benchmark the free energy profiles predicted by MLIPs against both classical MD results and experimental trends, evaluating their transferability and accuracy for complex interfacial systems.

Finally, we integrate the free energy profiles obtained from molecular simulations into a physics-informed continuum model of the electric double layer. This multiscale approach allows us to investigate how differences in atomistic parameterization propagate to macroscopic observables such as differential capacitance and the structure of the electric double layer. Through this combined methodology, we demonstrate the critical impact of force field and MLIP parameterization on the predicted behavior of ions at electrochemical interfaces and provide guidelines for the selection and refinement of interatomic potentials for interfacial modeling.
\section{\label{sec:methods}Computational Methods}

\setlength{\tabcolsep}{5pt}
\renewcommand{\arraystretch}{1.4}
\begin{table*}[ht]
\caption{12-6 Lennard-Jones force field parameters for the ions \ce{Na+}, \ce{Cl-} and \ce{F-} with charges +1.0, -1.0 and -1.0, respectively. The values are taken from the literature: Jorgensen from Ref.~\cite{Jensen2006}, Cheatham from Ref.~\cite{Joung2008}, Merz from Ref.~\cite{Sengupta2021} and Dang from Ref.~\cite{Dang1992,Smith1994}. In Refs.~\cite{Joung2008, Sengupta2021} the value $R_\text{min}$ is converted to $\sigma$ with the formula $\sigma = R_\text{min}/2^{1/6}$.}
\begin{tabular}{|l|cc|cc|cc|c|}
\hline
\multirow{2}{*}{Force field} & \multicolumn{2}{c|}{\ce{Na+} ($q=+1.00$)}                   & \multicolumn{2}{c|}{\ce{Cl-} ($q=-1.00$)}  & \multicolumn{2}{c|}{\ce{F-} ($q=-1.00$)}  & \multirow{2}{*}{Mixing rule}  \\ 
            &  $\varepsilon_{ii}$ [kcal/mol] & $\sigma_{ii}$ [\AA{}] & $\varepsilon_{ii}$ [kcal/mol] & $\sigma_{ii}$ [\AA{}] & $\varepsilon_{ii}$ [kcal/mol] &  $\sigma_{ii}$ [\AA{}] & \\ \hline \hline
Jorgensen   & 0.0005 & 4.0700    & 0.7100 & 4.0200    & 0.7100 & 3.0500  & Geometric         \\
Cheatham    & 0.0874 & 2.4393    & 0.0356 & 4.4777    & 0.0034 & 4.1035  & Lorentz-Berthelot \\
Merz        & 0.0276 & 2.5996    & 0.6380 & 4.0981    & 0.2414 & 3.2678  & Lorentz-Berthelot \\
Dang        & 0.1300 & 2.3500    & 0.1000 & 4.4500    & 0.2000 & 3.1680  & Lorentz-Berthelot \\ \hline

\end{tabular}
\label{tab:FF_params}
\end{table*}

\subsection{Classical Molecular Dynamics}
Classical MD simulations were run using LAMMPS~\cite{LAMMPS}, stable release 29 August 2024. To model the water, we used the TIP3P potential with modified charges and LJ parameters optimized for simulations using Ewald summation~\cite{Price2004} ($\sigma_\text{O} = 3.1880$, $\varepsilon_\text{O} = 0.10200$, $q_\text{O}=-0.830$, $q_\text{H}=0.415$, no \ce{H} LJ parameters), while the gold surface was modeled using the parameters from Heinz \textit{et al.}~\cite{Heinz2008} ($\sigma_\text{Au} = 2.6290$, $\varepsilon_\text{Au} = 5.29000)$. The force field parameters for the ions (\ce{Na+}, \ce{Cl-} and \ce{F-}) were taken from the literature shown in Tab.~\ref{tab:FF_params}. To see the dependence of the result on the choice of mixing rules to compute inter-particle coefficients, simulations with both the geometric mixing rule and the Lorentz-Berthelot mixing rule (arithmetic for $\sigma$, geometric for $\varepsilon$) were performed. Unless specified explicitly, the mixing rule is the same as the one used for the ion force field parameters shown in Tab.~\ref{tab:FF_params}. The timestep was set to 1.0~fs using the SHAKE algorithm~\cite{Ryckaert1977, Andersen1983} to constrain the \ce{O-H} bond and \ce{H-O-H} angle of water molecules. We used a cut-off of 12.0~\AA{} with a smooth energy switching function between 11.0~\AA{} and 12.0~\AA{} to treat the Lennard-Jones and Coulomb interactions and a particle-particle particle-mesh solver (PPPM) method with a grid spacing of $10^{-5}$ in $k$-space to treat long-range electrostatic interactions.

The system of study consists of an orthogonal cell elongated along the c-axis, with an aqueous electrolyte placed in contact with two gold \ce{Au(111)} surfaces. The initial configuration was obtained by generating a solvent box of size $a=28.85$~\AA{}, $b=29.98$~\AA{}, $c=70.00$~\AA{} with a single ionic species (either \ce{Na+}, \ce{Cl-} or \ce{F-}) solvated in 2,024 water molecules to reach a target water density of 1.0~g/cm$^3$. \red{The length of the unit cell was chosen to be sufficiently large such that the self-interaction of the charged ion with its periodic images is negligible.} The packing of the water molecules was initialized with the help of the software PACKMOL~\cite{Martinez2009}. Afterward, the solvent box was inserted next to a gold slab of 1440 atoms (12 atomic layers each). This resulted in a system of total size $a=28.85$~\AA{}, $b=29.98$~\AA{}, $c=99.27$~\AA{}, with the $c$ direction perpendicular to the \ce{Au(111)} metal surfaces.

The equilibration of the simulation box was achieved with a multiple-step procedure: the atomic positions were initially relaxed with a combination of the steepest descent and conjugate gradient algorithms with an energy tolerance of $10^{-6}$~kcal~mol$^{-1}$ and a force tolerance of $10^{-6}$~kcal~mol$^{-1}$ \AA$^{-1}$, respectively. The simulation box was then gradually heated up to 298~K in the NVT ensemble during 100,000 steps (100~ps) and the temperature was held constant for an additional 50,000 steps (50~ps). In both cases, we used a temperature coupling constant of 0.1~ps. The box size was subsequently equilibrated using an isothermal-isobaric NPT ensemble~\cite{Nose1984, Hoover1985} for 1~ns under an isotropic pressure of 1~atm and temperature coupling and pressure piston constant of 0.1~ps and 1.0~ps, respectively. Afterward, the box was deformed so that its size matched the average lattice parameters over the last 500~ps of NPT trajectory. This procedure set the final box size. An additional thermalization of 1.5~ns in an NVT ensemble was performed to further equilibrate the system before production runs.

Metadynamics calculations to estimate the free energy of the ions as a function of their distance along the z-axis from the gold surface were done using the COLVARS module implemented in LAMMPS~\cite{Fiorin2013}. The one-dimensional free energy profile was stored on a discrete grid with a spacing of 0.1~\AA{}. The biasing potential was obtained by adding repulsive Gaussian hills of weight 0.001~kcal/mol and width twice the grid spacing every 1,000 timesteps (1.0~ps). Each free energy profile was obtained by combining the contributions of 12 replicas (multiple walker metadynamics~\cite{Raiteri2006}) communicating every 15,000 timesteps (15~ps).

\red{
To test the effect of metallicity on the metadynamics results, we also conducted simulations with the constant potential method (CMP) as implemented in LAMMPS using the ELECTRODE package~\cite{Ahrens-Iwers2021, Ahrens-Iwers2022}. To achieve this, we performed simulations using the finite-field approach, which -- to impose a potential difference between the two electrodes bounding the cell -- uses an electric field across a periodic cell instead of non-periodic boundary conditions~\cite{Dufils2021,Tee2022}. We set the potential difference between the left and the right gold surfaces to 0~V, and the sum of the total charge on both electrodes was constrained to 0 to simulate the electrochemical interface under open-circuit, charge-neutral conditions. The reciprocal width of electrode charge smearing $\eta$ (with $\eta=1/\sigma$) was set to 1.805~\AA{}$^{-1}$ unless specified otherwise. When performing dynamics with the CPM, only the electrolyte atoms were time integrated, while the gold atoms were fixed for the entirety of the simulation.
}

\subsection{Machine Learning Molecular Dynamics}
Machine learning-enabled MD were performed within the ASE framework~\cite{ase-paper} by using the \texttt{FAIRChemCalculator}~\cite{UMA-paper, Levine2025} and the first published UMA-small model (UMA-S, version 1.0). First, we tested three different UMA tasks (OMat, OMol, and OMC) by comparing the ML-predicted energy to the DFT energy for a small system comprised of gold, water, and dissolved \ce{NaF}. Since UMA-S(OMat) performed better than the other tasks (Fig.~\ref{suppfig:bench_uma-s}), we decided to use it for the subsequent MD simulations.

To estimate the effective temperature at which to run the simulations, we decided to match the experimental water diffusivity at room temperature~\cite{Holz2000}. This was performed by equilibrating water boxes of size $20\times20\times20$~\AA{}$^3$ (267 water molecules) at different temperatures, ranging from 250~K to 525~K. The systems were first relaxed with 150 steps of the FIRE algorithm~\cite{Bitzek2006}, then the temperature was slowly increased up to the target value with a Langevin thermostat~\cite{Allen2017}. Finally, the MD boxes were equilibrated in the NPT Berendsen ensemble~\cite{Berendsen1984} under ambient pressure (1.01325~bar) and the same target temperature. The timestep employed was 0.5~fs. The compressibility constant used in the Berendsen barostat was $4.57\cdot10^{-5}$~bar$^{-1}$. The temperature coupling constant $\tau_T$ and pressure coupling constant $\tau_p$ were 100~fs and 1,000~fs, respectively. We ran trajectories of 60~ps and monitored the water density over time. The water diffusion coefficient was calculated using the Einstein relation:
\begin{equation}
    \left\langle \left| \mathbf{r}(t) - \mathbf{r}(0) \right|^2 \right\rangle = 2 n D t \enspace ,
\end{equation}
where $r(t)$ is the position of the oxygen atoms at the time $t$, $n$ is the dimensionality of motion, and $D$ is the diffusion coefficient. We choose a temperature of 383~K for the remaining ML MD simulations, as at this temperature the calculated diffusivity was close to the experimental one (see Fig.~\ref{suppfig:H2O_diff_coeff}). In particular, we observed that the lack of a van der Waals dispersion energy-correction term resulted in a lower than bulk water density and water diffusivity for UMA-S(OMat). The UMA-S(OMol) task -- while not capable of correctly predicting the energies and forces of the slab interfacial system -- did predict values for water diffusivity and water density that match experimental observables. Correcting the UMA-S(OMat) task with an additional DFT-D2~\cite{Grimme2006} or DFT-D3~\cite{Grimme2010} dispersion term did fix the low water density predicted by UMA-S(OMat). However, this resulted in a computational efficiency of more than an order of magnitude lower than the original task, which would have made the rest of this work impossible to perform in a reasonable timeframe. To increase the accuracy of UMA-S(OMat) at describing electrode/electrolyte interfaces, we suggest that, in the future, a van der Waals dispersion energy-correction term should be added, either by fine-tuning the available model or adding the correction on-the-fly. All results from the MD simulations of water are available in the Supplementary Information.

The ML free energy profiles of the ions in water as a function of their distance from the \ce{Au(111)} surface were measured with umbrella sampling~\cite{Roux1995}, as implemented in the ASE integration of the PLUMED package~\cite{Bonomi2009,Tribello2014,Bonomi2019,Tribello2024}. For the atomic interactions, we used the UMA-S model with the OMat task. The system was composed of a slab of 6 \ce{Au(111)} layers (288 atoms), that was previously relaxed, in contact with 432 water molecules and a single ionic species (\ce{Na+}, \ce{Cl-} or \ce{F-}) in an elongated box size of $a=17.64$~\AA{}, $b=20.36$~\AA{}, $c=51.40$~\AA{} to match the water density of 1~g/cm$^3$. First, the temperature was ramped up to 383~K over 70~ps with the help of a Langevin thermostat~\cite{Allen2017}. Subsequently, we performed multiple windows umbrella sampling, where in each window we placed a harmonic potential with spring constant $k=0.5$~eV/\AA{}$^2$ at 0.5~\AA{} intervals from the gold surface. Dynamics were run with a Bussi thermostat~\cite{Bussi2007} at 383~K using a timestep of 0.5~fs and a temperature coupling constant $\tau_T$ of 100~fs. We sampled the biased trajectories for a total of 500~ps at intervals of 25~fs, and finally, we calculated the potential of mean force by reweighting the distribution of the collective variable (\emph{viz.}, the perpendicular distance between the ion and the gold surface) using the weighted histogram method implemented in the WHAM program~\cite{WHAM}.

\subsection{Continuum Modeling}
Ionic density profiles as a function of the distance to the interface, potential of zero charge (PZC), and simulated differential capacitance were obtained using a one-dimensional continuum model~\cite{Baskin2017,Baskin2019,SanzMatias2024}. The continuum model is based on a modified Poisson-Boltzmann functional for the free energy:
\begin{equation}
    \label{eq:cmodel}
    \dfrac{F[\mathbf{n}(z)]}{A\,dz} = \mathbf{n}(z) \cdot \left( \boldsymbol{\mu}^0 + \mathbf{q} \, \varphi(z) + \boldsymbol{\Phi}(z) \right) - \ln W(\mathbf{n}(z)) \enspace ,
\end{equation}
written as a free energy density in units of thermal energy ($k_\text{B} T$). We use vector notation to include the spatial dependence of all relevant species': concentrations, $\mathbf{n}(z)$; bulk chemical potentials, $\boldsymbol{\mu}^0$; charges, $\mathbf{q}$; and adsorption potentials, $\boldsymbol{\Phi}(z)$. The mixing entropy at each position $z$, $\ln W(\mathbf{n}(z))$, accounts for steric effects, size disparity, and finite concentration of the species~\cite{DebyeHuckel_note}:%\footnote{Missing from this particular expression are terms related to ion-ion interactions, such as those of Debye-H\"uckel, Born, or the Mean Spherical Approximation.}:
\begin{align}
    -\ln W(\mathbf{n}(z)) ={}&
    \sum_i \lambda_i n_i(z) \ln \left(\lambda_i n_i(z) dV\right) \notag \\
    &{}+ n_h(z) \ln \left(n_h(z) dV\right) \enspace .
    \label{eq:lnWeff_V}
\end{align}
$\lambda_i$ is the relative volume of species $i$, whose volume, $V_i=\lambda_i dV$, is referenced to the minimum over all species (including the solvent), $dV = \min (\{V_i\}, V_{sol})$. $n_h(z)$ is defined as:
\begin{equation}
    n_h(z)=\frac{1}{dV}-\boldsymbol{\lambda} \cdot \mathbf{n}(z) \enspace.
\end{equation}
The local electrostatic potential, $\varphi(z)$, is computed as a solution to the one-dimensional Poisson equation: 
\begin{equation}
    \dfrac{d}{dz} \left [ \varepsilon_0 \varepsilon_r(z) \dfrac{d\varphi(z)}{dz}  \right ]= - \mathbf{q} \cdot \mathbf{n}(z) = -\rho(z) \enspace ,
\end{equation}
where $\varepsilon_0$ is the vacuum permittivity and $\rho(z)$ the $z$-dependent charge density profile. \red{The continuum model employs a half-cell setup; therefore, the reference potential is that of the electrolyte, and the local electrostatic potential is expressed as the difference in potential between the electrode and electrolyte.} To account for solvent structuring at the interface, we allow for a spatially dependent relative permittivity $\varepsilon_r(z)$ by modeling it with a smoothened step-like function:
\begin{equation}
    \label{eq:dielectric_profile}
    \varepsilon_r(z) = 1 + \frac{\varepsilon_\text{H$_2$O} - 1}{1 + \exp\left(-\alpha (z - z_0')\right)} \enspace,
\end{equation}
with $\alpha=5.0$~\AA{}$^{-1}$, $z_0'=0.8$~\AA{} and $\varepsilon_\text{H$_2$O}=80$~\cite{Baskin2019} (see Fig.~\ref{suppfig:H2O_eps_fun}).

The self-consistent minimization of the free energy functional with respect to $\mathbf{n}(z)$ and $\varphi(z)$ yields the thermodynamic equilibrium of species concentration and local electrostatic potential profile as a function of the distance to the electrode. The surface charge $\sigma_\text{int}$ on the electrode balances the charge accumulated nearby in the electrolyte and is obtained by integrating the charge distribution of the species and multiplying by $-1$:
\begin{equation}
    \label{eq:surface_charge}
    \sigma_\text{int} = -\int_0^\infty \rho(z)dz = -\int_0^\infty \mathbf{q} \cdot \mathbf{n}(z) dz\enspace.
\end{equation}
Each vector component, $q_i$, $n_i(z)$, represents the charge and the concentration of species $i$, respectively.
The differential capacitance $C_\text{diff}$ as a function of applied voltage $\varphi$ was obtained by the numerical differentiation:
\begin{equation}
    C_\text{diff}(\varphi) = \frac{\Delta \sigma_\text{int}(\varphi)}{\Delta\varphi(0)} \enspace.
\end{equation}
\red{The PZC is obtained using Eq.~\ref{eq:surface_charge} by iteratively finding the potential difference at which the continuum model predicts the surface charge of the electrode to be zero}. Our boundary conditions assume a perfect metal surface, which makes the PZC dependent only on the nature of the electrolyte (concentration and relative ion sizes) and its interaction with the electrode (adsorption potentials).

Input parameters to the continuum model were directly obtained from MD simulations. We used the species-specific volumes ($V_i$) of 125~\AA{}$^3$ for \ce{Na+}, 25~\AA{}$^3$ for \ce{F-}, and 50~\AA{}$^3$ for \ce{Cl-}. The volume for the sodium cation includes the presence of its first solvation shell and was calculated by averaging the excluded molecular volume (comprising either 5 or 6 water molecules) computed with the ARVO algorithm~\cite{Bua2005}. The volumes of \ce{F-} and \ce{Cl-} were computed assuming a sphere with a radius that is the mean of their ionic radii and the start of the corresponding first peak of the ion-O radial distribution function. The adsorption profiles $\boldsymbol{\Phi}(z)$ in Eq.~\ref{eq:cmodel} correspond to the free energy profiles computed with metadynamics or umbrella sampling. The curves were shifted along the $z$-axis by $-1.18$~\AA{} so that the origin $z=0$ corresponds to the jellium edge of Au(111)~\cite{Lang1973,Heinz2008}.

\subsection{Other Software and Methods}
The details of the algorithm for the analysis of the site-specific adsorption are explained in our previous publication~\cite{Roncoroni2023}. Notably, we used UMAP for dimensionality reduction~\cite{McInnes2018,McInnes2020,Healy2024}, HDBSCAN for hierarchical clustering~\cite{McInnes2017}, and FASTOVERLAP~\cite{Griffiths2017} and IRA~\cite{Gunde2021} for structure alignment.

For the manipulation, analysis, and visualization of atomic structures, we used the Atomic Simulation Environment (ASE)~\cite{Larsen2017} and the software OVITO~\cite{Stukowski2010}. All the plots in this article were generated with the help of Matplotlib~\cite{matplotlib} and seaborn~\cite{seaborn}.

%The experimental three-electrode setup consisted of an \ce{Au(111)} interface electrode in contact with a solution of NaF in pure water at the concentrations of 0.01~M, 0.1~M, and 1~M. The differential capacitance measurements were performed in a similar setup as in Ref.~\cite{Shatla2021}, by measuring the in- and out- of phase components of the interfacial impedance.

\section{\label{sec:results}Results and Discussion}

\subsection{\label{sec:results:subsec:CMD}Free Energy Profiles of Ions at the Au(111)/Water Interface from Classical MD}

\begin{figure*}[ht]
    \centering
    % First row
    \subfloat[\ce{Na+} parameters]{%
        \includegraphics[width=0.48\textwidth]{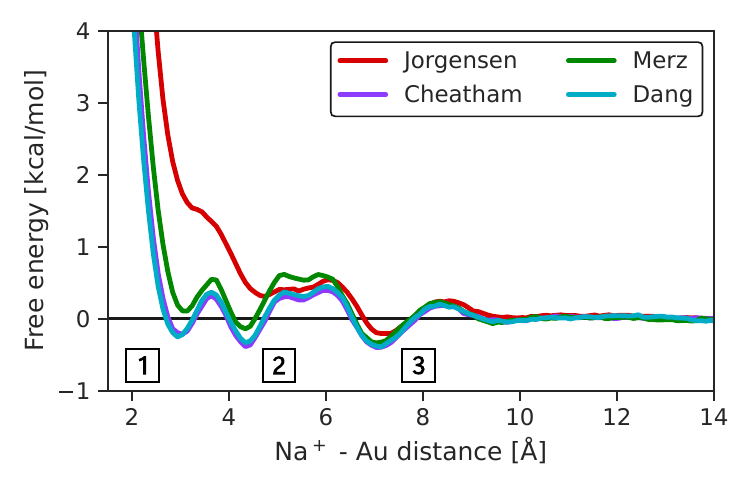}%
        \label{fig:Na_fe}%
    }
    \hfill
    \subfloat[\ce{F-} parameters]{%
        \includegraphics[width=0.48\textwidth]{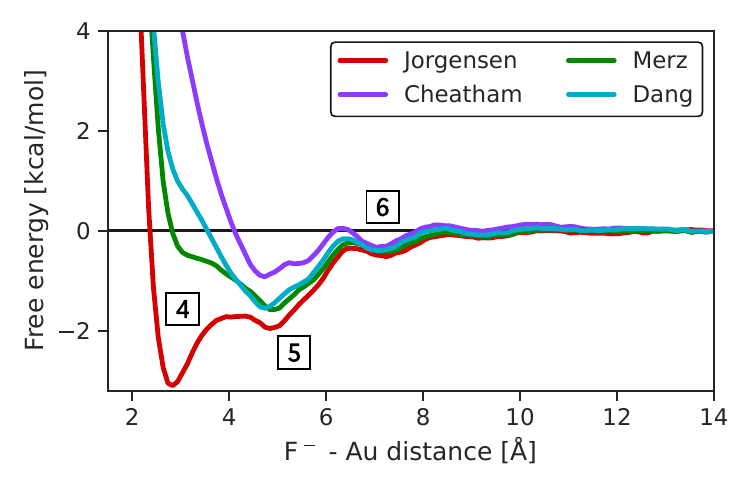}%
        \label{fig:F_fe}%
    }
    % Second row
    \par\medskip
    \subfloat[\ce{Cl-} parameters]{%
        \includegraphics[width=0.48\textwidth]{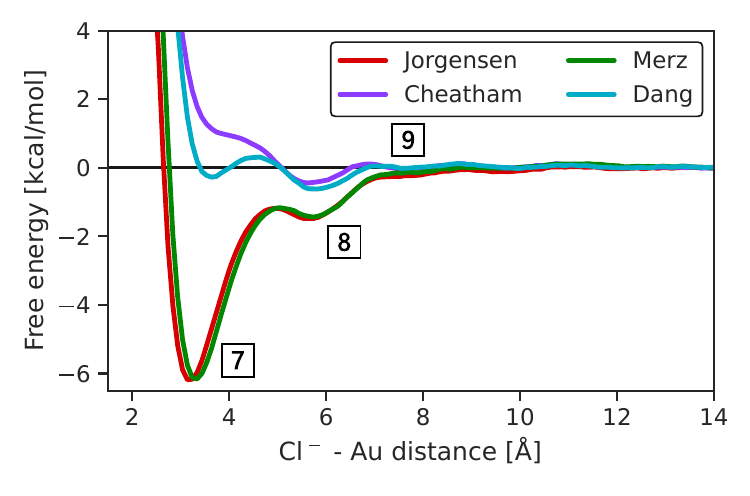}%
        \label{fig:Cl_fe}%
    }
    \hfill
    \subfloat[Schematic of the system with one ion (in this case \ce{Na+}.]{%
        \includegraphics[width=0.48\textwidth]{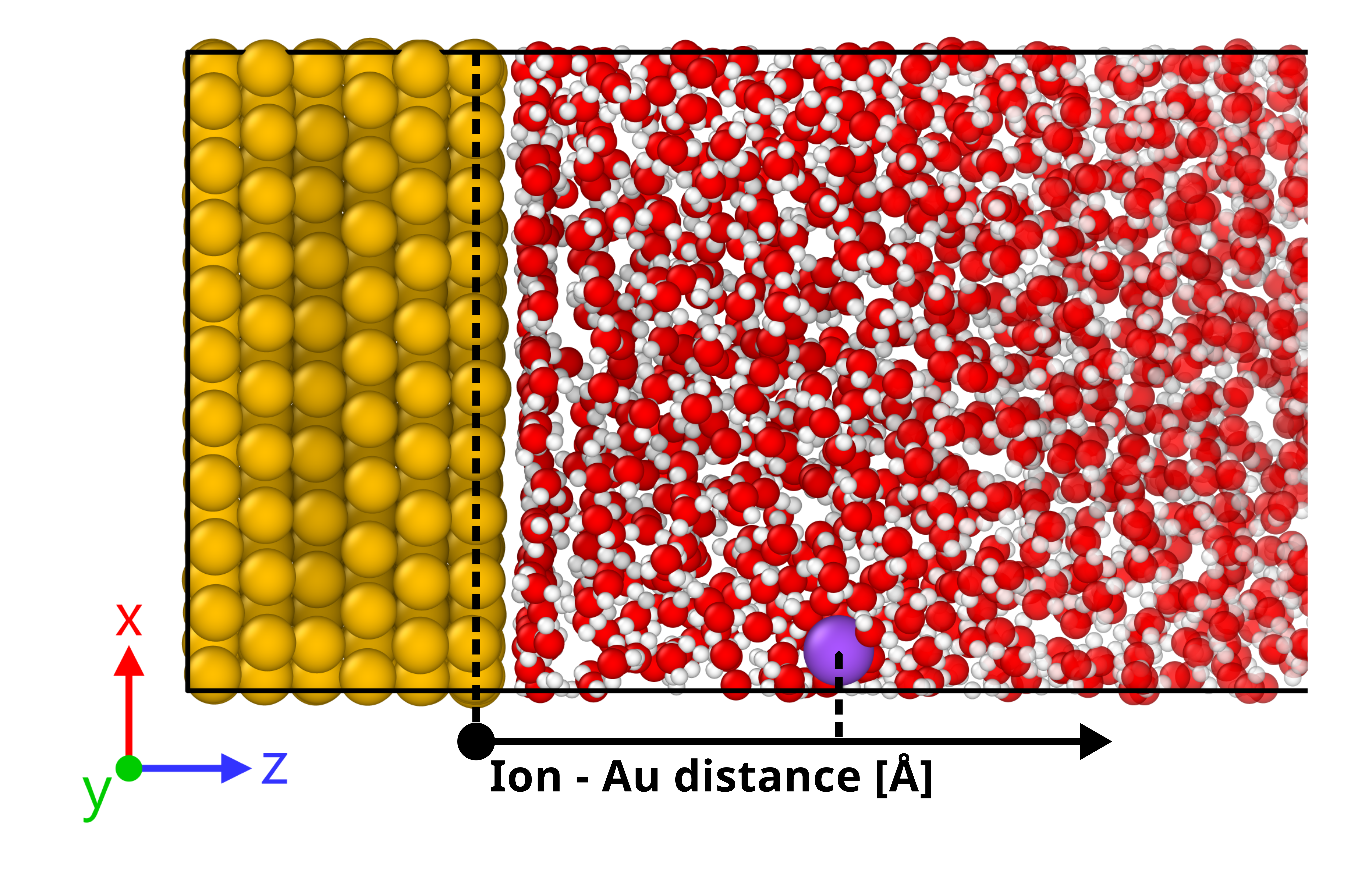}%
        \label{fig:system_fe}%
    }
    \caption{Comparison of the free energy profile of \ce{Na+} (a), \ce{F-} (b), \ce{Cl-} (c) in TIP3P water as a function of their distance from a \ce{Au(111)} surface. (d) Schematic of a classical MD system with a single \ce{Na+} ion, water, and the gold slab.}\label{fig:free_energies}
\end{figure*}

For each of the LJ parameter sets listed in Table~\ref{tab:FF_params}, we calculated the free energy profiles as a function of the distance to the water-gold interface using metadynamics for all three ions: \ce{Na+}, \ce{F-} and \ce{Cl-} (Fig.~\ref{fig:free_energies}). Ion distance is calculated as the perpendicular distance between the ion and the center of mass of the outermost gold layer. For all cases, the free energy curves were referenced to the bulk electrolyte by averaging the free energy between 14~\AA{} and 20~\AA{} and subtracting those values from the curves. We observe the effect of the interface on the free energy of the ion up to approximately 12~\AA{} from the surface, where it eventually converges to zero (with reference to the bulk electrolyte) for all models. Deviations from bulk behavior in the free energy are caused by the structuring of the water molecules at the interface and by the pairwise interactions between the gold surface and the ion.

In the case of \ce{Na+}, in Fig.~\ref{fig:Na_fe}, we do not observe deep local minima for any of the parameters studied, which indicates that there is no strong specific adsorption at the interface. Yet, the free energy profiles show a series of smaller local minima. The free energy calculated with the force field parameters of Cheatham, Merz, and Dang are very similar: the profiles are characterized by three local minima at 3.0~\AA{}, 4.3~\AA{}, and 7.1~\AA{}, respectively. Jorgensen's parameters, on the other hand, suggest a greater repulsion of the ion from the interface. In addition to the local minimum at 7.2~\AA{}, there is another minimum at 4.7~\AA{} that lies 0.3~kcal/mol above the free energy in the bulk.

For fluoride (\ce{F-}), in Fig.~\ref{fig:F_fe}, the free energy profiles of the four parameter sets reveal a wide range of behaviors for the ions at the surface. While all models predict a local minimum in the free energy at approximately 4.9~\AA{}, the depth and character of this minimum and the shape of the free energy at shorter Au-ion distances differ substantially. Cheatham's parameter set indicates predominantly repulsive interactions at shorter distances, with only a shallow minimum, suggesting that the fluoride anion does not spontaneously reach the gold surface. In contrast, simulations performed with Jorgensen's parameters result in a pronounced minimum of -3.1~kcal/mol at 2.8~\AA{} from the surface. This is indicative of spontaneous fluoride adsorption and direct contact with the interface. Merz and Dang parameterizations fall in between those two extremes: there is no clear additional energy minimum and, therefore, we would expect a weaker attraction of the fluoride ion to the interface.

Finally, the free energy profile for the \ce{Cl-} anion is shown in Fig.~\ref{fig:Cl_fe}. For chloride, the different force field parameters end up predicting two opposite behaviors at the interface. Jorgensen and Merz's parameters result in a clear global minimum located at 3.2~\AA{} from the interface, suggesting a strong ion-gold attraction of -6.0~kcal/mol. In contrast, Dang and Cheatham's parameters do not predict strong specific adsorption of the ion at the interface. In the case of Dang's parameters, there is a small local minimum at 3.6~\AA{} where the free energy is comparable to that in the bulk. Using Cheatham's parameters, we would actually predict repulsion of the ion as it approaches the gold.

\begin{figure}[ht]
    \centering
    \includegraphics[width=\columnwidth]{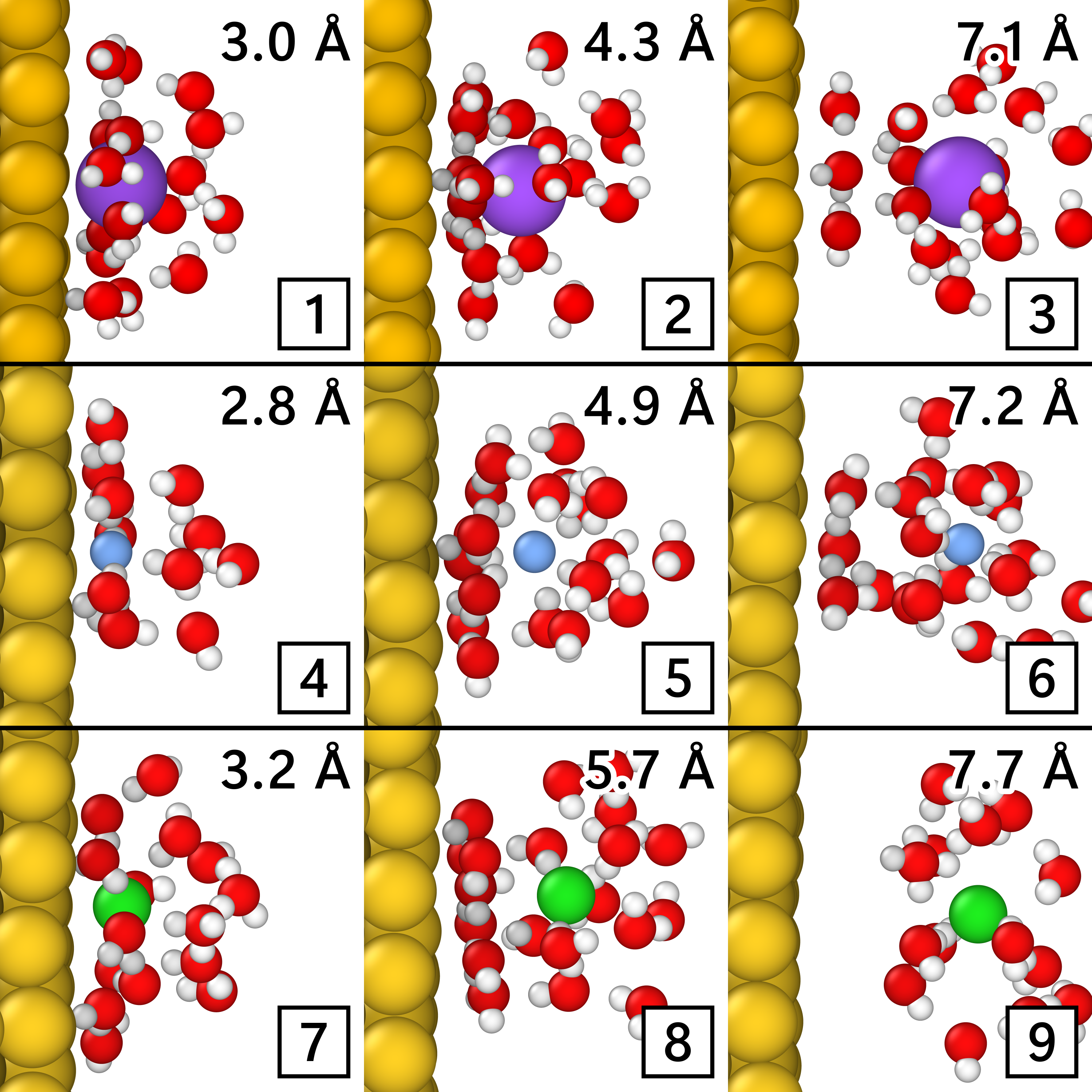}
    \caption{Snapshot from the classical MD trajectory for \ce{Na+} (purple, top, Cheatham), \ce{F-} (blue, center, Jorgensen), and \ce{Cl-} (green, bottom, Merz). The other elements are Au (yellow), O (red), and H (white). The distance from the \ce{Au(111)} surface for each snapshot is given in the top-right corner. Water molecules above a cutoff radius of 5~\AA{} from the ion were removed for visualization purposes. The numbering corresponds to the minima labeled in Fig.~\ref{fig:free_energies}.}
    \label{fig:FE_minima_rendering}
\end{figure}

Fig.~\ref{fig:FE_minima_rendering} presents snapshots from the classical MD trajectories, illustrating the configurations in which the ions occupy a local minimum in the free energy profiles shown in Fig.~\ref{fig:free_energies}. Notably, panels 1, 4, and 7 represent the scenario where the ions are specifically adsorbed to the gold surface and have partially shed their water solvation environment. Panels 2, 5, and 8 show the ions separated from the gold by one water layer, and the solvation shell of the ions is in contact with the gold. In panels 3, 6, and 9, the ions are completely solvated by water molecules. It is their solvation shell that interacts with the water molecules that are forming an adsorbed layer on the metal surface. In the case of chloride (panel 9), the water adlayer is not visible as it is outside the cutoff of 5~\AA{} from the ion. From the renderings, we can conclude that the minima observed in a range of $2.5-3.5$~\AA{} all feature specifically adsorbed ions at the gold surface.

The fact that the free energy curves of all parameter sets tested are comparable when the ion-gold distance is greater than approximately 5~\AA{} suggests that it is the gold-ion interaction that dominates the free energy at short range, while the shape of the free energy at larger distances is dictated by the ion-water interaction and how the solvation environment of the ion interacts with the layering of the solvent. Notably, the larger $\sigma_{ii}$ and smaller $\varepsilon_{ii}$ values of Jorgensen's sodium parameters are a probable cause of the increased repulsion between the ion and the metal surface. The same argument explains the behavior of fluoride with Cheatham's parameters, where the LJ parameter $\varepsilon_{ii}$ is approximately 20 times smaller than Jorgensen's. 

\begin{figure}[ht]
    \centering
    \includegraphics[width=\columnwidth]{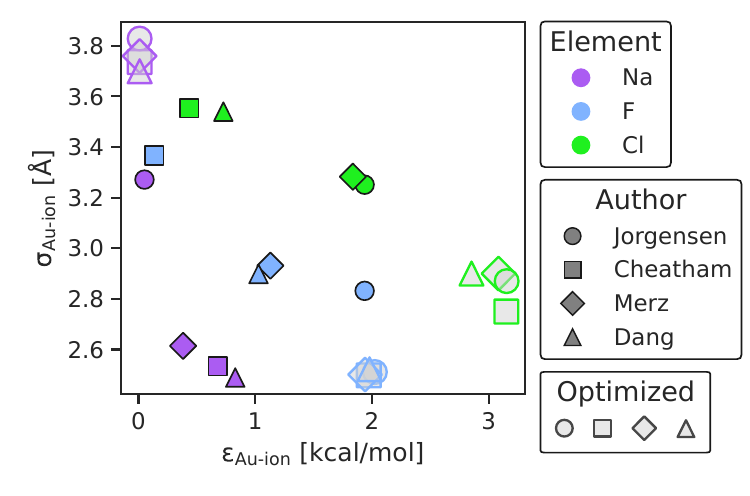}
    \caption{Effective force field parameters between the ions and the gold ($\varepsilon_{\text{Au-ion}}$ and $\sigma_{\text{Au-ion}}$) calculated with the mixing rules of Eq.~\ref{eq:Au-ion_mix}. The “optimized” parameters are the effective gold-ion interaction calculated after reparametrization as discussed in Sec.~\ref{subsec:cont_mod} and summarized in Tab.~\ref{tab:FF_params_opt}.}
    \label{fig:parameters_scatter}
\end{figure}

In Fig.~\ref{fig:parameters_scatter} we show the effective gold-ion Lennard-Jones interaction computed with the mixing rules specified in Table~\ref{tab:FF_params}:
\begin{align}
    \varepsilon_{\text{Au-ion}(\text{all})} &= \sqrt{\varepsilon_{\text{ion}}\cdot \varepsilon_{\text{Au}}} \notag \\[0.3em]
    \sigma_{\text{Au-ion}(\text{geometric})} &= \sqrt{\sigma_{\text{ion}}\cdot \sigma_{\text{Au}}} \notag \\
    \sigma_{\text{Au-ion}(\text{arithmetic})} &= \frac{1}{2} \left ( {\sigma_{\text{ion}} + \sigma_{\text{Au}}} \right ) \enspace. \label{eq:Au-ion_mix}
\end{align}
The positions of $\sigma_{\text{Au-ion}}$ and $\varepsilon_{\text{Au-ion}}$ on the plot support what is observed in the free energy curves; in particular, how three out of four models for sodium (Cheatham, Merz, and Dang) are very similar and cluster together, while the Jorgensen model deviates significantly. In the case of chloride, the models split into two distinct groups, with Cheatham and Dang forming one group and Merz and Jorgensen the other.

The differences observed in the free energy profiles for the three ions underscore the sensitivity of interfacial ion behavior to the choice of force field parameters. Previous studies have observed strong specific adsorption of \ce{Cl-} anions to the \ce{Au(111)} surface~\cite{Hamelin1978,Lipkowski1998}. At the same time, it is believed that -- while fluoride anions do specifically adsorb on gold  -- the strength of the adsorption is limited~\cite{Tucceri1985,Shatla2020,Adnan2024}. Indeed, when it comes to the strength of anionic specific adsorption at metal interfaces, studies have repeatedly shown the general trend \ce{F-} < \ce{Cl-} < \ce{Br-} < \ce{I-}~\cite{Tucceri1985,Markovic2002,Laurinavichyute2017,Shatla2020}. This argument is supported by the fact that larger anions tend to have a softer solvation shell that can more readily accommodate the presence of the metal. At the same time, larger anions also show increased polarization effects that lead to the formation of partial covalent character in the adsorption bond. Cation-specific adsorption has received comparatively less attention in the literature~\cite{Mills2014}. Although alkali metal cations are generally not considered to strongly specifically adsorb onto metal surfaces, there is evidence that they can influence the rate of oxygen reduction on the \ce{Au(111)} electrode~\cite{Strmcnik2011}. This effect is thought to arise from non-covalent interactions, rather than from specific adsorption or covalent bonding~\cite{Strmcnik2009, Singh2016}. Studies have shown that the strong solvation shell surrounding smaller cations makes it harder for the ion to exit the cage formed by the solvent, effectively increasing the energy needed to desolvate and adsorb to the metal surface. In contrast, larger cations, such as \ce{Cs+}, which possess a more weakly bound solvation shell, do exhibit a greater tendency toward surface adsorption.

In summary, based on our metadynamics calculations for the free energy of the ions, none of the force field parameters from a single author simultaneously satisfy the behavior predicted by experimental and computational studies. For sodium, the only force field parameter set that does not indicate specific adsorption at the gold surface is that of Jorgensen. For fluoride, Dang and Merz parameters are more consistent with the theory that the ion weakly specifically adsorbs to the surface. For chloride, only Jorgensen and Merz parameters suggest strong specific adsorption of this anion to the gold. From the comparison of the free energy between the four different force fields presented in the section above, we can conclude that large values of the LJ self-interaction term $\varepsilon_{ii} \geq 0.5$~kcal/mol result in strong ion adsorption, small values of $\varepsilon_{ii} \leq 0.05$~kcal/mol result in repulsion at the interface, and values in between result in mixed behavior with local minima dictated by solvent layering. 

We note that not all gold-ion interaction parameters were employed with the same mixing rules: Jorgensen's parameters assume geometric mixing, while Dang, Merz, and Cheatham assume Lorentz-Berthelot mixing. However, we performed metadynamics calculations with both the geometric and arithmetic mixing rules (as defined within LAMMPS), and the results indicate negligible impacts on the free energy profile (Supplementary Information~\ref{suppfig:FE_geo_ari}). Ultimately, these findings indicate that a separate optimization of the ion-gold LJ parameters, beyond simple mixing rules, is necessary to obtain consistent adsorption free energies for ions at the water-gold interface. 

\subsection{Influence of Lennard-Jones Parameters on Ion Adsorption Free Energies}
\label{subsec:influence_lj_params}

\begin{figure}[ht]
    \centering
    \subfloat[Adsorption free energy of \ce{Na+} as a function of \ce{Na+-Au} $\varepsilon$.]{
        \includegraphics[width=\columnwidth]{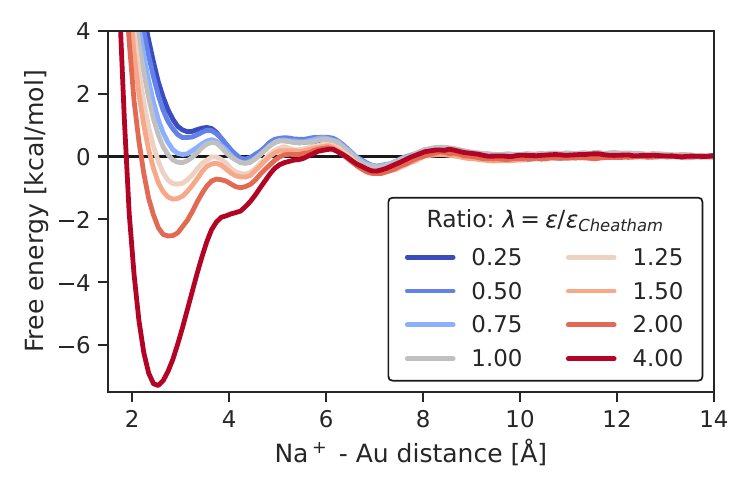}
        \label{fig:FE_epsilon}
    }
    \hfill
    \subfloat[Free energy difference between the curves in (a) with 12-6 LJ fit.]{
        \includegraphics[width=\columnwidth]{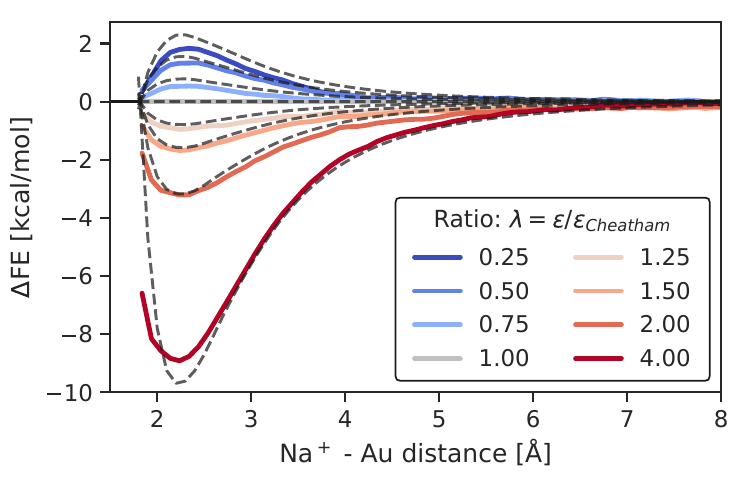}
        \label{fig:dFE_epsilon}
    }
    \caption{(a) Free energy profile of a \ce{Na+} cation calculated with metadynamics while varying the ratio of the LJ $\varepsilon$ parameter with respect to that computed by the geometric mean (Cheatham's parameters). (b) Free energy difference between the curves shown in (a).}
    \label{fig:FE_combined}
\end{figure}

An alternative approach to modifying the ion-gold interaction is to explicitly set the LJ cross terms -- $\sigma_{ij}$ and $\varepsilon_{ij}$ -- rather than calculating them with mixing rules. We performed metadynamics for the force-field parameters of Cheatham for \ce{Na+} while explicitly tuning the parameter $\varepsilon_{\text{Au-Na}^+}$ for the interaction between the gold and the sodium cation. Under normal circumstances, the value would be computed using the Lorentz-Berthelot mixing rules of Eq.~\ref{eq:arithmetic_mr}, as the geometric mean of the gold and \ce{Na+} parameters. Note that this change only affects the explicit ion-metal LJ term, and we do not modify the interaction with the solvent to avoid changing the solvation behavior of the ion in the bulk electrolyte. In Fig.~\ref{fig:FE_epsilon} we show the effect on the free energy that we obtain by varying $\varepsilon_{\text{Au-Na}^+}$. We define the parameter:
\begin{equation}
    \lambda=\dfrac{\varepsilon_{\text{Au-Na}^+}}{\varepsilon_{\text{Au-Na}^+(\text{geometric})}}\enspace,
\end{equation}
which is the ratio between the explicit ion-gold interaction and that calculated as the geometric mean. By tuning the value of the $\lambda$ parameter, it is possible to either make the ion more attracted to the interface (higher $\lambda$) or more repelled by it (lower $\lambda$).

As discussed above, the effect of the metal-ion interaction is seen up to a distance of 6~\AA{} from the surface as further proof that the minima at 7~\AA{} can be mostly attributed to the layering of the solvent. In Fig.~\ref{fig:dFE_epsilon}, we show the difference between the free energies with respect to the geometric mean $\varepsilon_{\text{Au-Na}^+(\text{geometric})}$ (defined as $\lambda=1.00$ in the plot). The dashed lines represent fits to the free energy difference that is calculated with a correction to the 12-6 LJ interaction between the ion and the gold atoms. We define $V_{\text{Au-ion}}$ as the metal slab-ion pairwise interaction energy:
\begin{equation}
    \label{eq:LJ_correction}
    V_{\text{Au-ion}}(z;\,\varepsilon,\sigma) = \sum_{i=1}^{N_{\mathrm{Au}}} V_{\mathrm{LJ}}\big(r_{ik}(z);\, \varepsilon,\, \sigma\big)\enspace, %\cdot \Theta(r_c - r_{ik}(z))\enspace,
\end{equation}
which is the sum of the LJ interactions between the ion and the gold atoms in the slab calculated from Eq.~\ref{eq:LJ_potential} as a function of the distance of the ion from the surface $z$. $r_{ik}$ is the distance between the ion and gold atom $i$. The LJ term is technically also a function of the explicit gold-ion interaction $\sigma_{\text{Au-Na}^+}$ and $\varepsilon_{\text{Au-Na}^+}$, and by changing $\lambda$ we are altering the effective potential acting on the ion when it is close to the surface. Properly correcting the free energy for the presence of the altered LJ interaction would require a full reweighting of the MD trajectory under this new bias. However, as a coarse first-order approximation, we assume that the effect of the interaction term $\varepsilon_{\text{Au-Na}^+}$ only contributes to differences in enthalpy and specifically only to the non-bonded ion-gold interaction. Therefore, we define $\Delta V$ as the 12-6 LJ correction term:
\begin{equation}
    \Delta V(z;\,\varepsilon,\sigma) =  V_{\text{Au-ion}}(z,\varepsilon,\sigma) - V_{\text{Au-ion}}(z,\varepsilon_{\text{ref}},\sigma_{\text{ref}})\enspace,
\end{equation}
which is the difference between the metal slab-ion pairwise interaction energy with the new values of $\varepsilon$ and $\sigma$ with respect to the reference ($\varepsilon_\text{ref}$ and $\sigma_\text{ref}$). The correction is averaged across the plane parallel to the gold surface with a Boltzmann factor to account for the energy of the configurations. This is particularly relevant at small perpendicular distances, where configurations with the ion very close to gold atoms are unlikely to happen because of the strong repulsion.

The additional LJ term can properly predict the difference in the adsorption free energy for varying values of $\lambda$. The difference between the fit and the actual free energy differences deviates at higher values of $\lambda$ and at short ion-gold distance $z$. Indeed, site-specific adsorption plays an important role when the interaction between the gold surface and the ion is stronger, and we expect the \ce{Na+} ion to preferentially adsorb at specific sites on the \ce{Au(111)} surface. Therefore, the enthalpic approximation that does not account for entropic effects and the time-dependent accessibility of interstitial sites next to the gold plane may fail to accurately capture the true free energy landscape at high interaction strengths.

This simple relationship between the free energy profile for the adsorption of the ion and the LJ parameter $\varepsilon_{\text{Au-Na}^+}$ suggests that the re-parametrization of the force field terms can be done straightforwardly. The effect of changing the ion-metal interaction can be approximated without the need to perform additional metadynamics calculations, thus potentially saving substantial computational time and allowing for an iterative fit of the ion-gold interaction.

\begin{figure}[ht]
    \centering
    \includegraphics[width=\columnwidth]{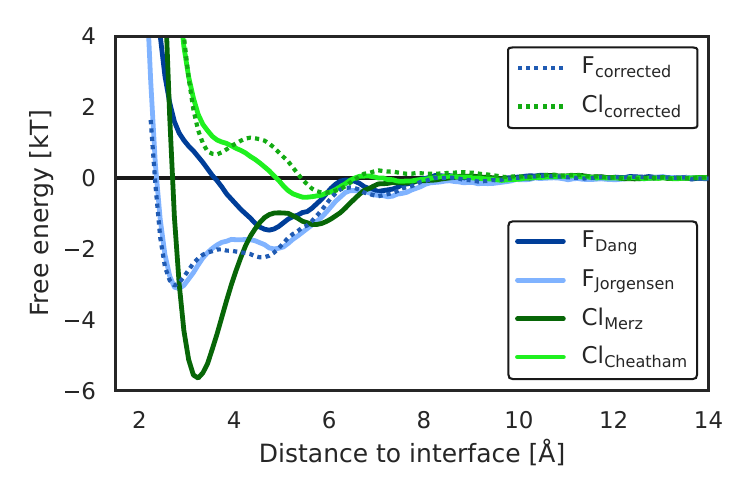}
    \caption{Lennard-Jones correction of the adsorption free energy for two sets of parameters. The dotted lines represent the estimated free energy when correcting from F$_{\text{Dang}}$ $\rightarrow$ F$_{\text{Jorgensen}}$ and from Cl$_{\text{Merz}}$ $\rightarrow$ Cl$_{\text{Cheatham}}$.}
    \label{fig:classical_LJ_correction}
\end{figure}

In Fig.~\ref{fig:classical_LJ_correction} we show how the correction term can be used to convert adsorption free energy between different classical force field parameters. By adding an extra attractive LJ potential, we can correct the fluoride adsorption profile of Dang (F$_{\text{Dang}}$) to better match that from Jorgensen (F$_{\text{Jorgensen}}$. At the same time, by adding a repulsive LJ interaction, we can transform Merz's chloride profile to better match Cheatam's. Because we are assuming that all the contributions except that from ion-gold remain the same, the quality of the fit is not as accurate as when we compare free energies within the same force field, as done in Fig.~\ref{fig:dFE_epsilon}. Yet, the overall trend is maintained.

In our simple case, the adsorption free energy can be represented as a function of five distinct contributions of individual pairwise interactions: water-water, water-gold, ion-water, ion-gold, and gold-gold. We explicitly ignore the ion-ion interaction, as we model the dilute limit with a single ionic species. In our comparison of different classical force field parameters for the ions, the gold-gold, water-water, and water-gold interactions remain unchanged. The gold-gold interaction defines the structure of the metal, including the lattice constant, the roughness of the surface exposed to the electrolyte, and the availability of interstitial sites for species to specifically adsorb (e.g., hollow and bridge sites). The water-water interaction defines the properties of bulk water, such as the formation of the hydrogen bond network. Those two components, together with the water-metal interaction, define the structure and the dynamics of water at the interface. Experimental and computational evidence support the presence of a characteristic double peak observed in the density profile of water $\rho_{\text{H$_2$O}}(z)$ and the presence of a chemisorbed water adlayer on the surface~\cite{Carrasco2012,Le2017,VelascoVelez2017,Geada2018,Gim2019,Clabaut2020,Darby2022,Wang2025}. This inherent solvent structure depends on the nature of the metal and acts as a template for the adsorption behavior of the ion. 

The structured layers of water at the metal interface play a central role in mediating ion adsorption~\cite{Willard2009}. The presence of organized water adlayers introduces both energetic and entropic barriers that ions must overcome to reach specific adsorption sites on the metal surface. In particular, the first water layer can compete with the ion and hinder direct contact at the surface~\cite{Dewan2014}. As a consequence, the adsorption process frequently involves partial or complete desolvation of the ion, as well as local restructuring of the interfacial water network. The extent to which an ion can displace interfacial water molecules or, alternatively, become stabilized within the existing solvent structure depends on its size, charge, and specific ion-water and ion-metal interactions~\cite{Serva2021b}.

In this study, the water-water and water-gold interactions are kept the same across all classical MD sampling -- we use the modified TIP3P parameters for water and Heinz's parameters for the gold. Only the ion-water and ion-gold LJ parameters are different. From our observation, the largest effect on ionic adsorption can be attributed to the ion-gold interaction, while the ion-water interaction has a limited effect. This observation is consistent with the fact that the force field parameters for ions dissolved in bulk water are optimized against experimental and quantum mechanical observables. Therefore, we should expect comparable interactions across the four force fields tested. In this regard, we cannot exclude the possibility that multiple combinations of $\varepsilon$ and $\sigma$ could lead to very similar electrolyte dynamics. While this assumption seems to hold for the behavior of the ions in bulk water, the same does not apply to the interfacial region.

\subsection{\red{Impact of Electrode Metallicity on Interfacial Ion Adsorption}}

\begin{figure}[ht]
    \centering
    \subfloat[Free energy]{
        \includegraphics[width=\columnwidth]{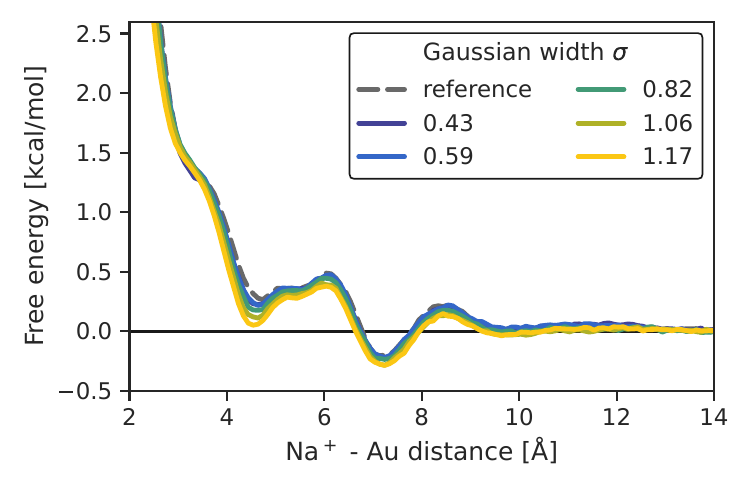}
        \label{fig:FE_ELE_eta}
    }
    \hfill
    \subfloat[Free energy difference]{
        \includegraphics[width=\columnwidth]{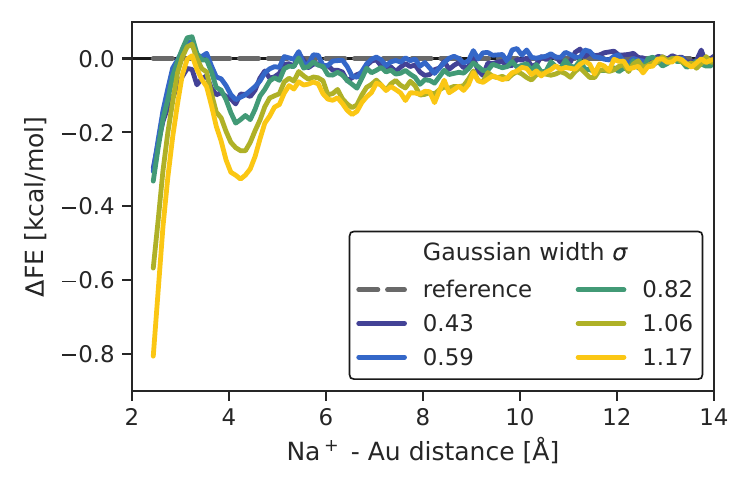}
        \label{fig:dE_ELE_eta}
    }
    \caption{(a) Free energy profile of a \ce{Na+} cation calculated with metadynamics while varying the Gaussian width $\sigma=1/\eta$ parameter (describing electrode charge smearing) compared to Jorgensen's parameters~\cite{Jorgensen1996}. (b) Free energy difference between the curves shown in (a).}
    \label{fig:ELECTRODE}
\end{figure}

\red{
Figure~\ref{fig:ELECTRODE} shows the effect that varying the Gaussian width $\sigma$ of the charge distribution has on the adsorption free energy of the sodium cation. The reference profile is from Jorgensen~\cite{Jorgensen1996} (see Fig.~\ref{fig:Na_fe}), where the gold atoms are modeled as neutrally charged Lennard-Jones spheres. The introduction of the CPM method does not noticeably change the overall shape of the adsorption free-energy profile, yet we observe increasing differences as the ion approaches the interface. Notably, a higher Gaussian width $\sigma$ correlates with a stronger adsorption at very short distances. This is consistent with the fact that a lower value of $\sigma$ is used to represent poor metals, and by increasing its value, the additional interaction between the metal and the ion due to the image charge effect increases.
}

\red{
The study of the effect of metallicity on ion adsorption at the interface is not new: in previous work from Serva \textit{et al.} -- which inspired this portion of our study -- it was already proven how changing the Gaussian width of the charge distribution on the electrode could be used as a proxy to tune the metallicity of the electrode~\cite{Serva2021}. The authors claim that an increase in the Gaussian width $\sigma$ is correlated with a noticeably higher adsorption of \ce{Na+} -- and to a lesser extent \ce{Cl-} -- at the \ce{Au(100)} interface. They also observed a progressive shift in the position of the adsorbed \ce{Na+} toward the gold while increasing $\sigma$ and even the appearance of inner sphere adsorbed \ce{Na+} located within the hollow sites of the metal surface. While the two computational setups are not directly comparable because of the different orientation of the metal surface exposed to the electrolyte (\ce{Au(111)} in this work vs. \ce{Au(100)} in Ref.~\cite{Serva2021}), the overall conclusion remains the same: increasing the Gaussian width $\sigma$ leads to stronger ionic adsorption, specifically close to the metal surface. We expect this behavior to be even more pronounced at finite concentration, where the ability of the metal to screen the presence of adsorbed charged species should facilitate packing of ions at the interface. It is worth noting that -- according to Fig.~\ref{fig:FE_ELE_eta} -- under no bias we would not expect to find a considerable population of species adsorbed at the metal surface. Indeed, even with a lowering of the adsorption free energy because of the CPM, the free energy difference between the bulk electrolyte and the surface would be about 3.0~kcal/mol ($\approx5.01$~kT at 300~K). A key difference between this work and Ref.~\cite{Serva2021} is the force field parameters for the ion: whereas Figure~\ref{fig:FE_ELE_eta} shows calculations with Jorgensen's parameters~\cite{Jorgensen1996}, the cited work used Dang's parameters~\cite{Dang1992,Smith1994}. According to our previous calculations comparing the different force fields, there is a substantial difference in free energy profile between Dang and Jorgensen, and Dang's \ce{Na+} parameters suggest the presence of a minimum close to the \ce{Au(111)} surface that is not observed in Fig.~\ref{fig:FE_ELE_eta}. Additionally, while we employed a modified TIP3P water model, the authors relied on another water model specifically tuned for the water-gold interface~\cite{Berg2017}.
}

\subsection{Ion Adsorption Free Energies from Machine-Learned Interatomic Potentials}

Here, we present how we can rely on newly developed MLIPs to compute the free energy of ions at solid interfaces. We focus on the recently published UMA potential because of its generic applicability to different systems across many applications. By performing umbrella sampling of ions at a gold surface, we are testing the model's limits in describing a complex system such as the electrochemical interface. 

\begin{figure}[ht]
    \centering
    \subfloat[Umbrella sampling with UMA-S(OMat).]{
        \includegraphics[width=\columnwidth]{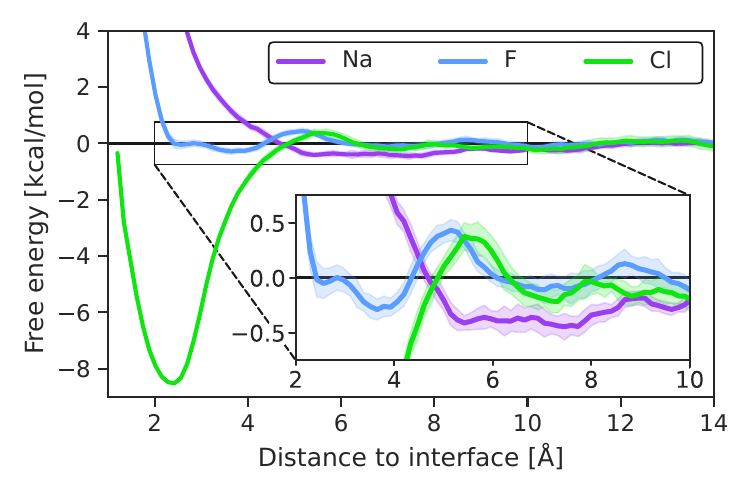}
        \label{fig:FE_UMA_OMat}
    }
    \hfill
    \subfloat[Water density with UMA-S(OMat).]{
        \includegraphics[width=\columnwidth]{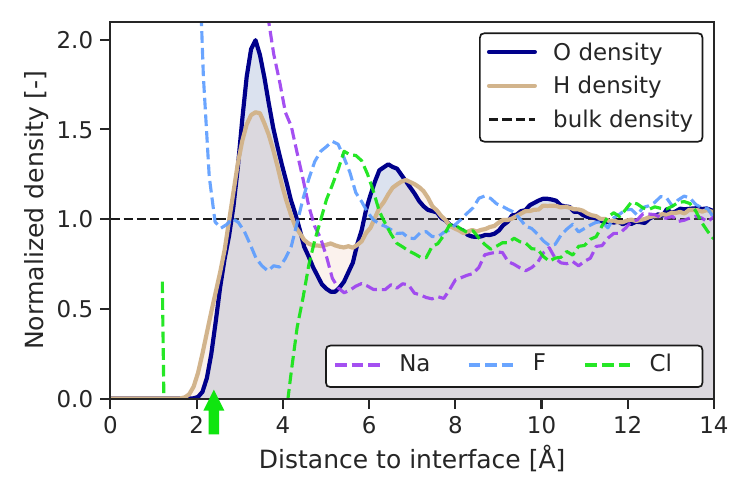}
        \label{fig:UMA_rho}
    }
    \caption{Molecular dynamics using the UMA-S(OMat) ML interatomic potential: (a) Comparison of the free energy of \ce{Na+}, \ce{F-}, and \ce{Cl-} ions as a function of their distance from a \ce{Au(111)} surface calculated with umbrella sampling. (b) z-dependent density profile of water (oxygen and hydrogen) at the \ce{Au(111)} interface. In (b), the free energies from (a) are overlaid as a guide to the reader.}
    \label{fig:UMA_FE_rho}
\end{figure}

The free energy profile for the three ions calculated using umbrella sampling and the UMA-S(OMat) MLIP is shown in Fig.~\ref{fig:FE_UMA_OMat}. The inset of the graph shows a close up of the free energy 2~\AA{} to 10~\AA{} from the interface. The shaded regions represent the uncertainty of the free energy calculated with bootstrapping error analysis.

The difference in the free energy between the two anions is clear: UMA-S(OMat) predicts a strong specific adsorption of \ce{Cl-} (-8.40~kcal/mol at 2.3~\AA{} from the interface), while for \ce{F-} there is not a noticeable energy gain when approaching the gold surface, apart from a small local minimum of -0.30~kcal/mol at 3.6~\AA{}. The large energy gain for chloride suggests that the anion will tend to desolvate and stick to the gold surface. The strong repulsion for fluoride starts at 2.5~\AA{} from the center of mass of the outermost \ce{Au(111)} layer, indicating that the \ce{F-} anion as well can be in direct contact with the metal (this can still be considered specific adsorption, even if there is no net energy gain). The sodium cation is even less prone than fluoride to approach the surface: the free energy profile shows that there is a small energy gain between 4.8~\AA{} and 9~\AA{} from the interface. However, a strong repulsive barrier prevents the cation from desolvating and specifically adsorbing to the metal surface. The predictions from the UMA-S(OMat) potential are comparable to what is expected from the literature discussed in subsection~\ref{sec:results:subsec:CMD}.

\begin{figure}[ht]
    \centering
    \includegraphics[width=\columnwidth]{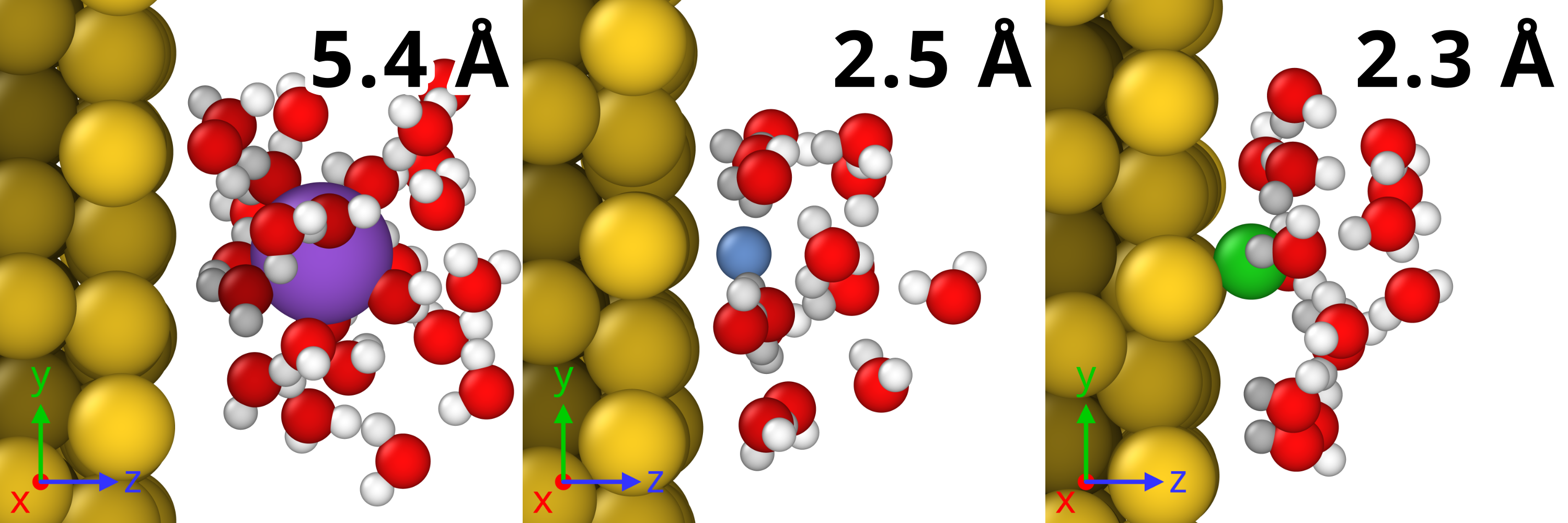}
    \caption{Snapshot from the MLMD trajectory for \ce{Na+} (purple, left), \ce{F-} (blue, center), and \ce{Cl-} (green, right). The other elements are Au (yellow), O (red), and H (white). The distance from the \ce{Au(111)} surface is shown for each snapshot in the top-right corner. Water molecules above a cutoff radius of 5.4~\AA{} from the ion were removed for visualization purposes.}
    \label{fig:MLMP_snapshots}
\end{figure}

In Fig.~\ref{fig:UMA_rho} we show the z-dependent water density profile calculated with the UMA-S(OMat) potential. We observe three peaks of decreasing intensity as we move away from the gold interface: a first peak of highly structured water between 2.0~\AA{} and 5.0~\AA{}, a second broader peak up to 8.7~\AA{} and a smaller peak up to 11.0~\AA{}. The structuring of the water molecules at the interface modulates to some extent the free energies from Fig.~\ref{fig:FE_UMA_OMat}. The minimum in the free energy for \ce{Na+} shows some correlation with the oxygen density of water, suggesting that the positive sodium cation prefers regions where there are many water-oxygen atoms that can solvate it. Yet, this behavior deviates as the ion approaches the gold, and the sodium free energy increases gradually across the first layer of adsorbed water. This suggests an aversion of the cation to shedding its solvation shell and indicates that, without applied bias, the \ce{Na+} ion does not specifically adsorb to the gold. It is rather its solvation shell that is in contact with the metal. In the case of fluoride, the local maxima in the free energy correlate with the presence of regions with a lower water density. We see a first minimum located within the first water layer, which means that both fluoride and water have a comparable distance from the gold.

The minimum in the free energy for chloride (green) at 2.3~\AA{} (outside the graph's bounds, green arrow) is located where the tail of the first water peak reaches 0. \ce{Cl-} can more readily approach the metal interface and occupy an intermediate region between the water molecules and the gold atoms. A rendering of the ions on the gold surface with their immediate solvation shell within a cutoff radius of 5.4~\AA{} is shown in Fig.~\ref{fig:MLMP_snapshots}.

\begin{figure}[ht]
    \centering
    \includegraphics[width=\columnwidth]{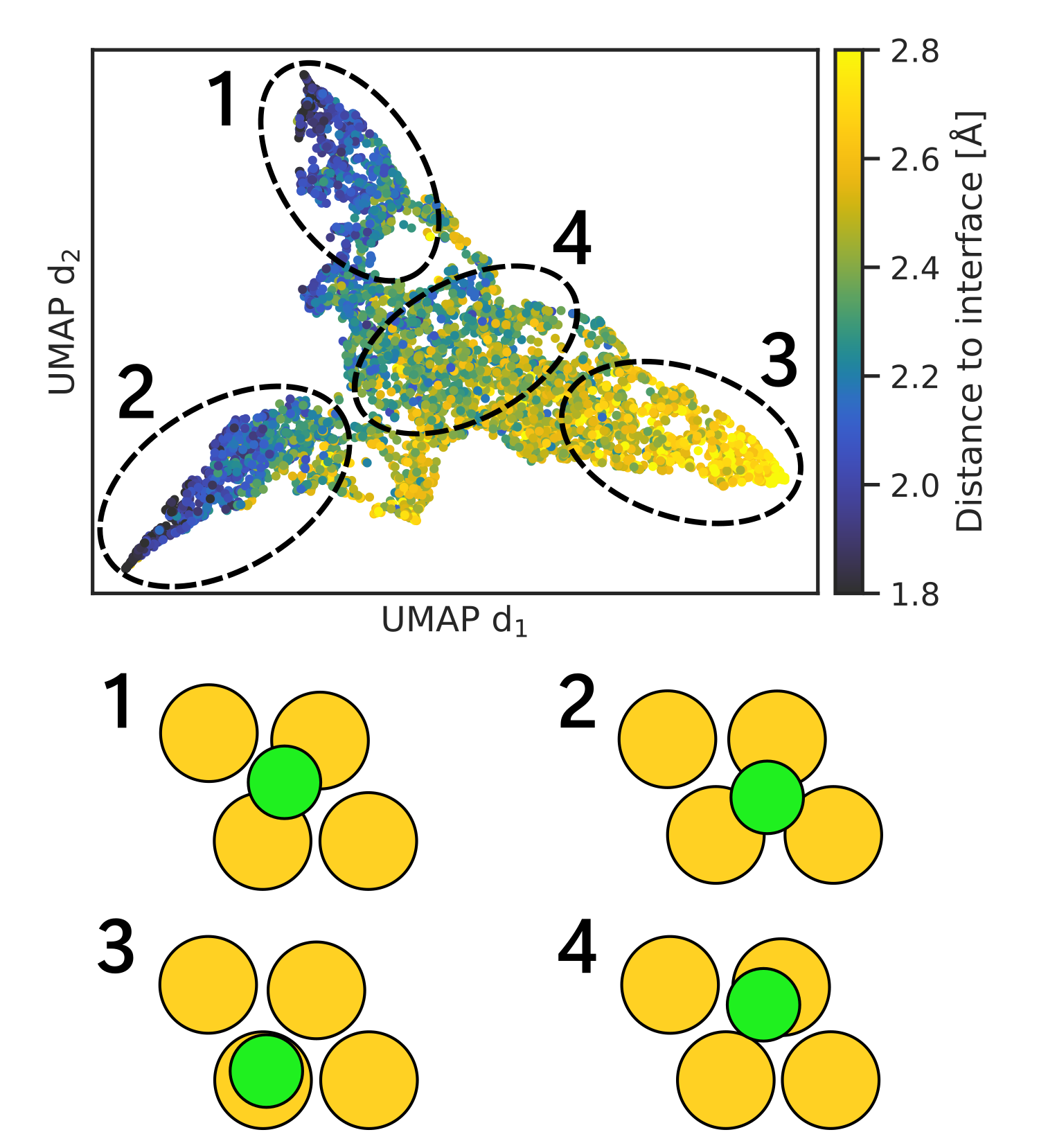}
    \caption{Top: UMAP projection of an MLMD trajectory of chloride adsorbed onto gold. Symbol color corresponds to the z-component of the distance vector between the gold interface and the ion. Bottom: representative structures showing the adsorption site of \ce{Cl-} (green) on gold (yellow) of the four regions circled on the UMAP space.}
    \label{fig:UMAP_Cl_adsorbed}
\end{figure}

To better understand the site affinity of the specific adsorption of chloride on the Au surface, we project the local coordination environment of \ce{Cl-} onto a two-dimensional manifold. Fig.~\ref{fig:UMAP_Cl_adsorbed} shows the UMAP projection of an MD trajectory using the UMA-S(OMat) potential, sampled when the anion is adsorbed to the gold surface. Each point on the plot represents a frame of the trajectory that has been colored according to the z-component of the distance vector between the gold interface and the ion. The triangle shape of the two-dimensional manifold suggests that there are three limiting cases: a bridge configuration with the ion shared between two neighboring gold atoms (1), a hollow configuration with the ion sitting in the space between three gold atoms (2), and a top configuration with the ion sitting predominantly above a single gold atom (3). The region denoted as (4) consists of transition structures with a configuration between the top and bridge limiting cases.

\subsection{Effect of Specific Ion Adsorption on the Continuum Modeling of the Electric Double Layer}
\label{subsec:cont_mod}

So far, we have focused on the effect of specific adsorption in the extremely dilute limit -- a single ion in a simulation box. To estimate the effect of the different force fields on macroscopic observable quantities -- and to bridge the gap between simulations and experiments -- we employ a one-dimensional continuum model to calculate the species density profile as a function of the distance from the gold interface under applied potential difference~\cite{Baskin2017,Baskin2019,SanzMatias2024}. The functional form of the continuum model is shown in Eq.~\ref{eq:cmodel}. The single-ion adsorption potential $\Phi$ obtained with enhanced sampling is directly used as input to the continuum model and accounts for ionic specific adsorption effects at the metal interface. We define two electrolyte systems: \ce{NaF} and \ce{NaCl}. Both salts are dissolved in \ce{H2O} in a $1:1$ ratio to ensure charge neutrality of the system.

\begin{figure}[ht]
    \centering
    \includegraphics[width=\columnwidth]{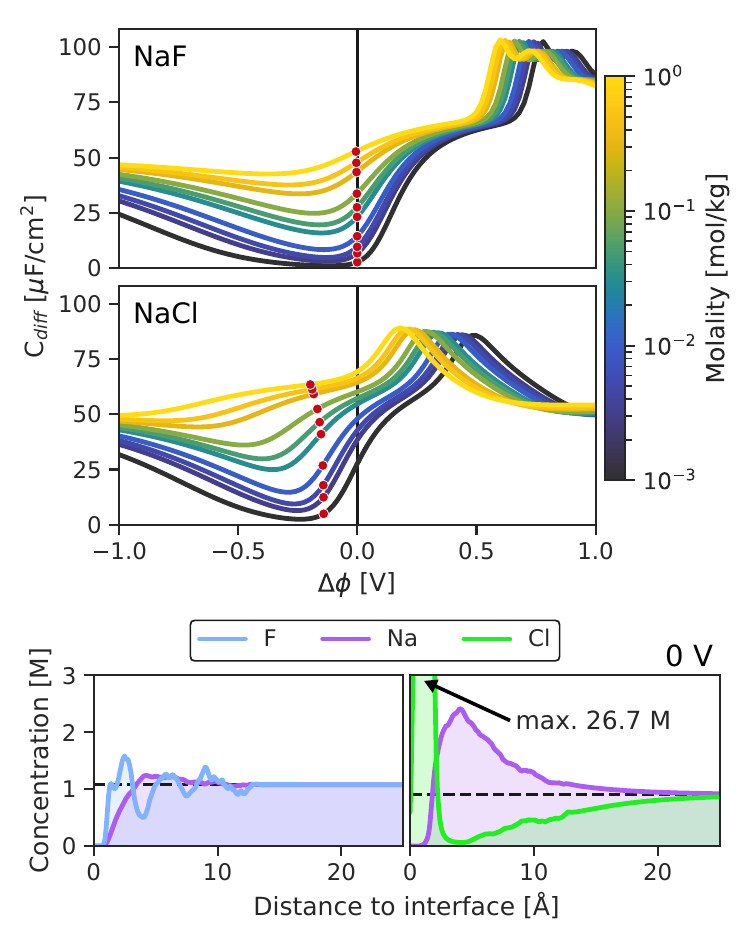}
    \caption{Top: differential capacitance plots as a function of molality for \ce{NaF} and \ce{NaCl} predicted with the UMA-S(OMat) MLIP. The position of the potential of zero charge (PZC) is highlighted in red. Bottom: z-dependent concentration profile as a function of the distance to the gold interface for 1~M solutions of \ce{NaF} and \ce{NaCl} without potential bias (0~V).}
    \label{fig:CM_Cd_OMat}
\end{figure}

We first investigate the results obtained with the ML potential. The adsorption behavior of UMA-S(OMat) represents two limiting cases: a system with two species that do not show strong specific adsorption at the metal interface (\ce{NaF}), and another where one of the two ions (\ce{Cl-}) does specifically adsorb (\ce{NaCl}). Fig.~\ref{fig:CM_Cd_OMat} (top) shows the differential capacitance as a function of electrolyte concentration predicted by the continuum model using the UMA-S(OMat) adsorption potentials. The position of the potential of zero charge (PZC) is highlighted in red, which is close to 0~V for \ce{NaF} and ranges between -0.15~V and -0.20~V for \ce{NaCl}. A shift in the PZC to larger magnitudes is usually dictated by the presence of strong specific adsorption of an ionic species~\cite{Vargas2018}. Indeed, to reach a neutral charge balance at the metal surface, the adsorbed ions need to be removed by applying a potential of the same sign as their charge. The values for $C_{\text{diff}}$ at larger positive and negative biases -- which, admittedly, would be outside the capacitive regime for water -- converge to the same value across a wide range of concentrations. In those boundary cases, the surface charge density is predominantly bound above by the maximal packing of species, determined by their effective sizes rather than bulk concentrations. Interestingly, in both cases, we notice how our continuum model predicts the PZC with an offset from the capacitance minima. This deviation from the simpler Gouy-Chapman-Stern description of the electric double layer~\cite{Shatla2021}, is due to specific adsorption and/or size asymmetry of the species, included in our modified Poisson-Boltzmann free energy density (Eq.~\ref{eq:cmodel}).

\red{Direct comparison between the differential capacitance curves predicted by the continuum model and experimental results requires careful consideration of several factors. In particular, the continuum model references the potential with respect to the electrolyte (half-cell system), while experimental setups often reference to a standatd electrode (for example the standard hydrogen electrode (SHE)). Additionally, there is a distinction between the capacitance values predicted by the thermodynamic model and scan-rate-dependent experimental measurements. Nevertheless, the capacitance values predicted by the continuum model and the characteristic camel-shaped curve are in reasonable agreement with experimental data reported for Au(111) in dilute solutions of weakly adsorbing anions~\cite{Kolb1986,Chao1989,Wang1992,Eberhardt1996,Pajkossy1996,Kerner2002,Smetanin2008,Adnan2024}.}

The lower half of the figure shows the concentration profiles of 1~M solutions of \ce{NaF} and \ce{NaCl} without applied potential. The strong specific adsorption of chloride is evident, with a maximum concentration close to the gold surface -- that lies well outside the limits of the plot -- at about 26.7~M, corresponding to a surface coverage of 94.6\%. To counteract the presence of negative ions at the surface, the concentration of \ce{Na+} is also higher in proximity to the chloride peak, forming a diffuse second ionic layer at the interface. By contrast, the $z$-dependent concentration profile for the two ions in the \ce{NaF} system is more uniform, and we observe only small deviations in concentration from the bulk electrolyte value.

\begin{figure*}[ht]
    \centering
    \includegraphics[width=\textwidth]{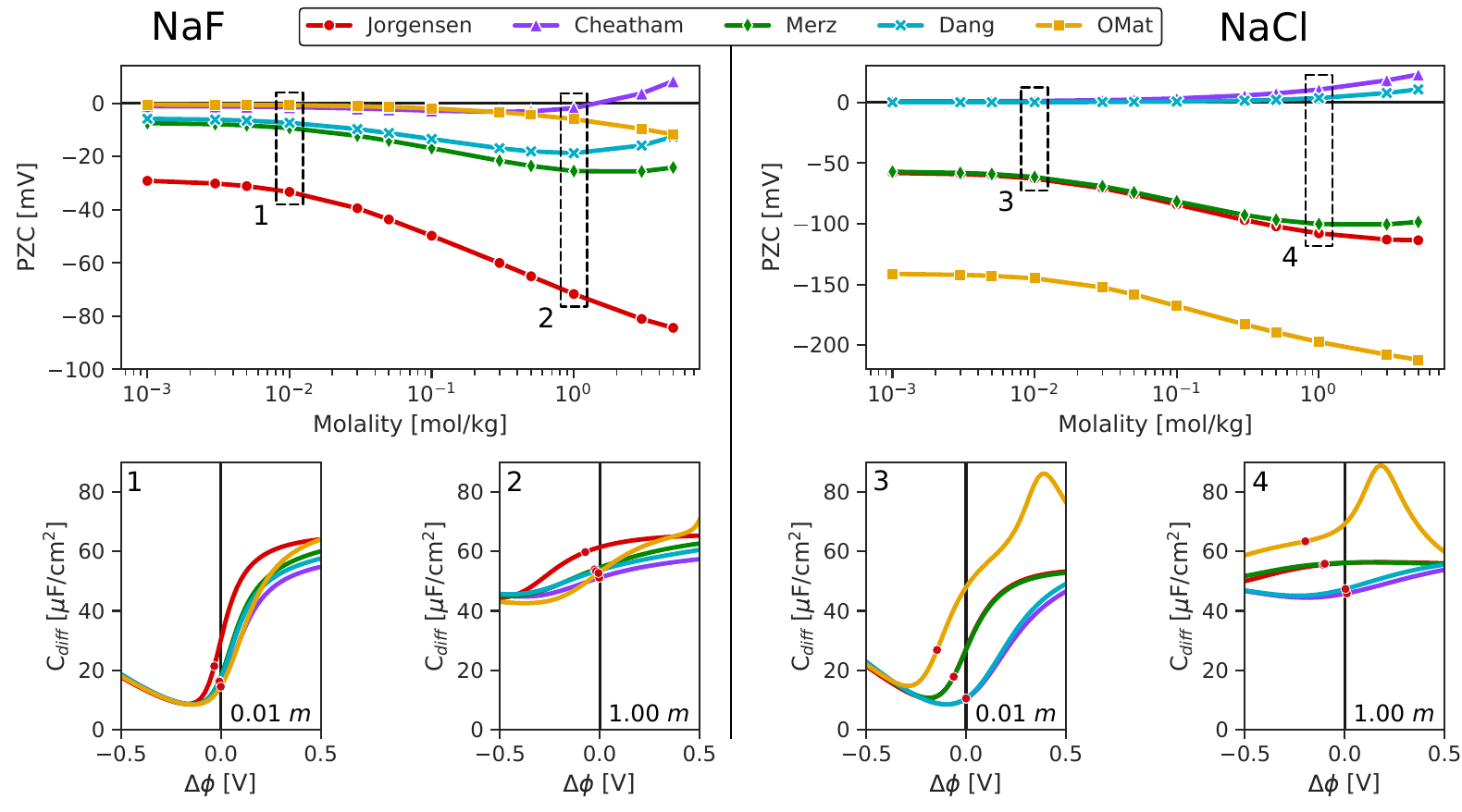}
    \caption{Comparison of macroscopic observables estimated with the continuum model when the classical free energies are sampled using standard mixing rules (geometric and Lorentz-Berthelot). Top: potential of zero charge (PZC) for \ce{NaF} (left) and \ce{NaCl} (right) systems as a function of molality. Bottom: differential capacitance as a function of the potential difference $\Delta\varphi$ for the insets of the top images. The position of the PZC on the $C_{\text{diff}}$ curves is identified by red symbols.}
    \label{fig:PZCs_combined}
\end{figure*}

In Fig.~\ref{fig:PZCs_combined} (top), we compare the effect of the adsorption free energy from the different force fields on the PZC as a function of molality for the two salts. The stronger specific adsorption of \ce{F-} explains the larger negative values of the PZC predicted using Jorgensen's parameters. Indeed, according to the free energy profile of the \ce{F-} anion, we expect a strong fluoride presence at the metal surface when there is no external bias applied. The same trend is observed by comparing the other force field parameters -- particularly for the \ce{NaCl} system -- where the correlation between specific adsorption and a shift in the PZC is evident. Interestingly, in the case of Cheatham (and Dang for \ce{NaCl}), we observe a positive PZC that increases with molality. Cheatham's anions show some degree of repulsion from the gold, while the sodium cation has a flat free energy profile that allows the \ce{Na+} to more readily approach the surface, resulting in a positive PZC. The insets in the lower part of Fig.~\ref{fig:PZCs_combined} show the differential capacitance as a function of applied potential for molalities of $0.01$~mol/kg and $1.00$~mol/kg respectively. We notice a correlation between the shape of the PZC curves and the differential capacitance; notably, Cheatham and Dang parameters predict very similar behavior across the board. Comparing the results from the classical MD with those obtained with the UMA-S(OMat) potential, we observe how the classical predictions severely underestimate both the PZC and the shape of the differential capacitance curve for the \ce{NaCl} system. Even with force field parameters that lead to strong chloride-specific adsorption (Jorgensen and Merz), the PZC is off by about 100~mV. For \ce{NaF}, apart from Jorgensen, the results tend to be closer to the MLIP data because of the overall weakly adsorbing behavior of the species involved.

We propose to correct the effective gold-ion interaction with the methodology described in Section~\ref{subsec:influence_lj_params}. We apply the fitting procedure to retrieve which set of LJ parameters for the effective gold-ion interaction results in a free energy profile similar to that obtained with the UMA-S(OMat) potential. Here, we use the ML as a surrogate model with higher accuracy than classical MD. This procedure can be extended to other levels of theory or more refined ML potentials. Tab.~\ref{tab:FF_params_opt} shows the optimized classical force field parameters for the four different classical potentials investigated in this work. The parameters are very similar as confirmation that the water-ion interaction -- which was not tuned -- is comparable. 

\setlength{\tabcolsep}{2.5pt}
\renewcommand{\arraystretch}{1.4}
\begin{table}[ht]
\caption{Explicit ion-gold interaction for the different force fields obtained with the fitting scheme to best reproduce the free energy sampled with the UMA-S(OMat) potential. The units for $\varepsilon_{\text{Au-ion}}$ are in kcal/mol and for $\sigma_{\text{Au-ion}}$ in \AA{}.}
\begin{tabular}{|l|cc|cc|cc|}
\hline
\multirow{2}{*}{Force field} & \multicolumn{2}{c|}{\ce{Na+}} & \multicolumn{2}{c|}{\ce{Cl-}}  & \multicolumn{2}{c|}{\ce{F-}} \\ 
&  $\varepsilon_{\text{Au-ion}}$ & $\sigma_{\text{Au-ion}}$ & $\varepsilon_{\text{Au-ion}}$  & $\sigma_{\text{Au-ion}}$ & $\varepsilon_{\text{Au-ion}}$ &  $\sigma_{\text{Au-ion}}$ \\ \hline \hline
Jorgensen   &  0.010  &  3.828  &  3.155  &  2.870  &  2.026  &  2.511  \\
Cheatham    &  0.010  &  3.735  &  3.152  &  2.750  &  1.973  &  2.498  \\
Merz        &  0.010  &  3.759  &  3.085  &  2.900  &  1.942  &  2.501  \\
Dang        &  0.010  &  3.699  &  2.854  &  2.900  &  1.980  &  2.522  \\ \hline
\end{tabular}
\label{tab:FF_params_opt}
\end{table}

The overall trend that emerges from the optimization is the very low $\varepsilon_{\text{Au-ion}}$ and increased $\sigma_{\text{Au-ion}}$ for \ce{Na+} to better reproduce the non-desolvating nature of the ion at the Au surface. For \ce{F-}, $\sigma_{\text{Au-ion}}$ is reduced to allow for the ion to better reach the gold surface; at the same time, $\varepsilon_{\text{Au-ion}}$ had to be increased so that the ion could compete with water and maintain a flat free energy profile. For \ce{Cl-}, shifting the local minimum to match the UMA-S(OMat) value at 2.3~\AA{} from the interface proved to be difficult, as the high non-linearity with respect to $\sigma$ of the LJ interaction makes the optimization procedure unstable. Yet, we increase the $\varepsilon_{\text{Au-ion}}$ to mimic the strong specific adsorption at the gold and match the adsorption energy of -8.40~kcal/mol at the interface. In general, we observed that fine-tuning $\varepsilon_{\text{Au-ion}}$ is more straightforward than $\sigma_{\text{Au-ion}}$. To overcome the lack of free energy data at small $z$ values (close to the gold), which are much less likely to be sampled during metadynamics, we find it effective to first manually lower $\sigma_{\text{Au-ion}}$ and then correct the free energy with the fitting procedure described above. Due to the high repulsion given by the LJ potential at short distances~\cite{Chiu2010}, we find that -- barring a change in analytic form of the potential -- a full reparametrization of the chloride-gold interaction cannot match the shape of the ML free energy\red{, and the simulation is unable to reproduce the adsorption types shown in Fig.~\ref{fig:UMAP_Cl_adsorbed}}. For \ce{F-} and \ce{Na+}, we can better fit the classical free energy to the ML free energy, yet we also observe in this case how the increase in repulsion at short distances for the UMA-S(OMat) is much less steep than for the classical models. The limitations of the widely used 12-6 LJ potential to describe interface interactions have already been reported in the literature for water~\cite{Berg2017} and benzene~\cite{Johnston2011}. Previous studies have also shown how the choice of water model can affect the structure of solvent at the interface and ion adsorption distances~\cite{Dix2018,Elliott2020}. This is another argument that highlights the importance of the force field choice and its parameters when describing interfacial systems.

\begin{figure*}[ht]
    \centering
    \includegraphics[width=\textwidth]{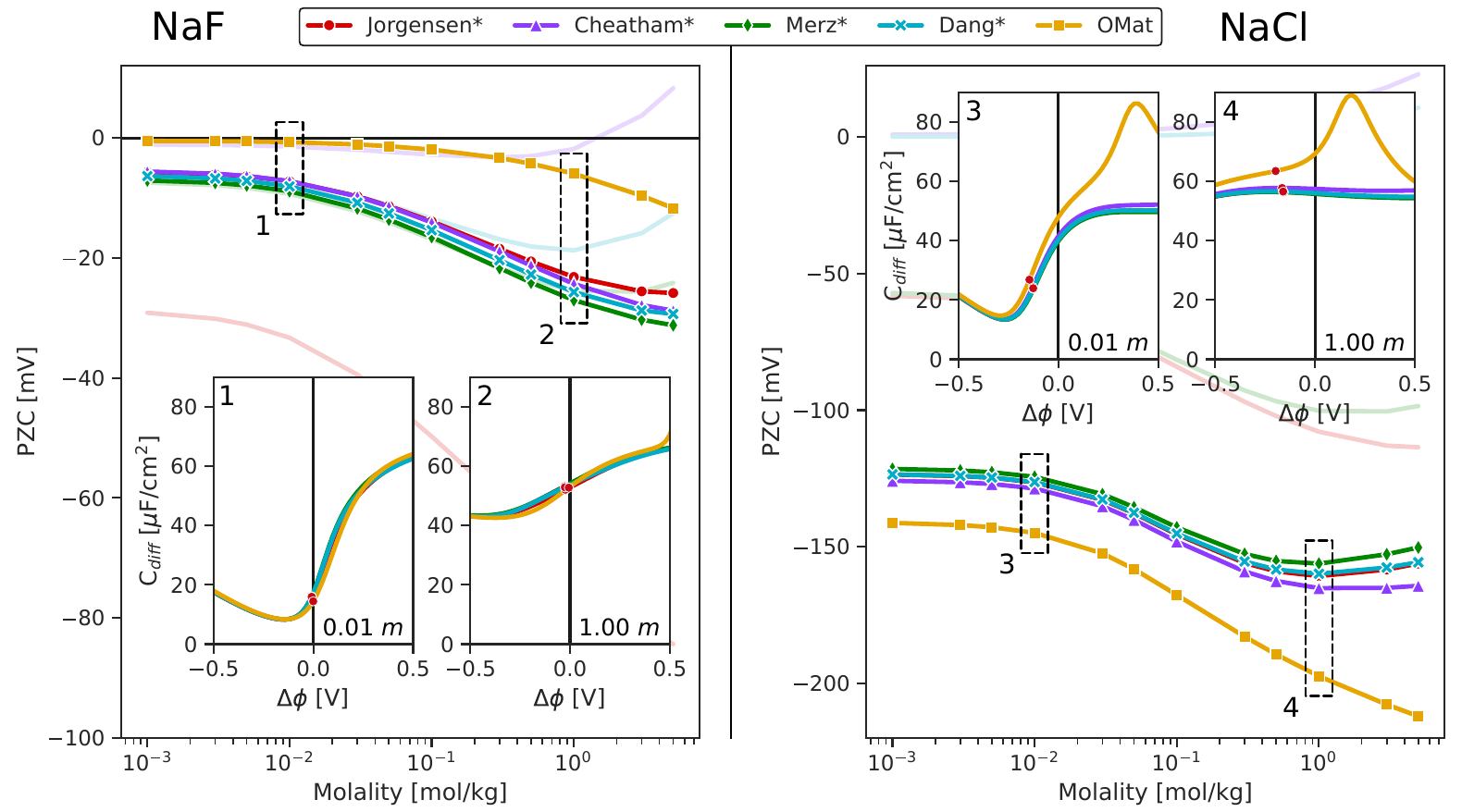}
    \caption{Comparison of macroscopic observables estimated with the continuum model when the classical free energies are sampled using the effective ion-gold interaction from Tab.~\ref{tab:FF_params_opt}. Main plot: potential of zero charge (PZC) for \ce{NaF} (left) and \ce{NaCl} (right) systems as a function of molality. The PZCs estimated with the standard mixing rules are added as semi-transparent lines as a guide to the reader. Insets: differential capacitance as a function of the potential difference $\Delta\varphi$ for the insets of the top images. The position of the PZC on the $C_{\text{diff}}$ curves is identified by red symbols.}
    \label{fig:PZCs_combined_optimized}
\end{figure*}

In Fig.~\ref{fig:PZCs_combined_optimized}, we show how the reparametrization of the gold-ion interaction affects the estimation of the PZC and the differential capacitance calculated by the continuum model. As expected, for all classical force fields, we obtain very similar PZCs trends and overlapping $C_{\text{diff}}$ curves. Small variations are still observed because of the different ion-water interactions. Those deviations, however, have minimal impact on the calculated PZC. The corrected free energies better reproduce the UMA-S(OMat) PZC curve and lie close to the target value. Particularly for the case of \ce{NaF}, where the PZC values deviate only up to 20~mV from each other. For the two concentrations plotted (insets 1 and 2), the shapes of the differential capacitance curves are in good agreement. Differences between the MLIP result and the classical force fields persist because of the intrinsic difference in the free energy dictated by the solvent layering and specific ion-water interactions. In the case of \ce{NaCl} we also observe a better match of the corrected classical PZC curves with those of the UMA-S(OMat). Notably, both Cheatham and Dang free energy profiles -- which predicted a positive PZC under standard mixing rules -- now show a similar trend to the other force fields. The values for the PZCs still fall short of that expected with the MLIP. This is mainly due to the different position of the minimum in the free energy of \ce{Cl-}, which lies closer to the interface for UMA-S(OMat). In particular, the effect can be observed while comparing the differential capacitance plots from insets 3 and 4, which show a deviation of the UMA-S(OMat) curve at positive potentials. Because of the step-like dielectric profile used in the continuum model, shown in Eq.~\ref{eq:dielectric_profile}, the region closest to the interface has a lower dielectric screening strength than in the bulk (see Supplementary Information, Fig.~\ref{suppfig:H2O_eps_fun}). The lack of solvent screening results in a higher electrostatic energy cost to accumulate of ions at the interface. Yet, if the ability of the anion to approach the interface outweighs the increase in electrostatic energy associated with the reduction in local dielectric constant, the net effect is an increased capacitance. The UMA-S(OMat) adsorption free energy allows chloride anions to more readily reach the gold-water interface where there is lower screening, effectively leading to a sharper potential drop in this narrow region.

\begin{figure}[ht]
    \centering
    \includegraphics[width=\columnwidth]{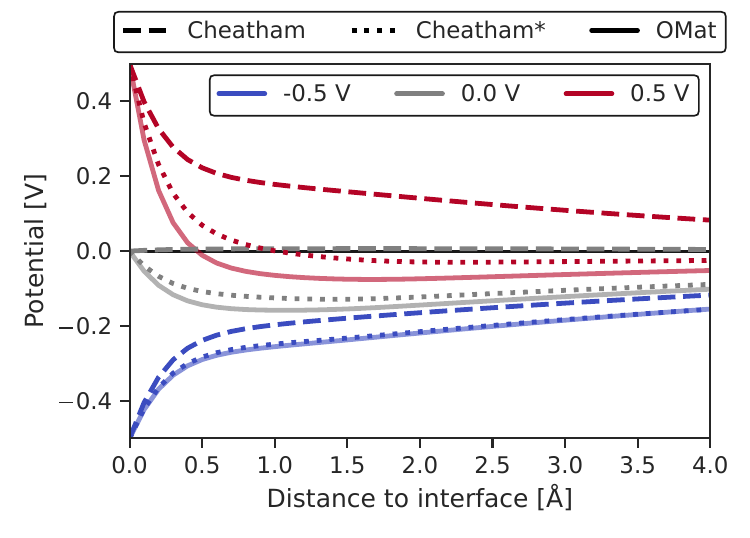}
    \caption{Local electrostatic potential profile $\varphi(z)$ of the \ce{NaCl} system (1.0~mol/kg) obtained from the continuum model for different potential differences between the interface and the bulk. The lines correspond to the result obtained with Cheatham under standard mixing rules (dashed lines), Cheatham with reparametrized gold-ion interaction (dotted lines), and the UMA-S(OMat) potential (solid lines).}
    \label{fig:dV_cheatham_omat}
\end{figure}

Fig.~\ref{fig:dV_cheatham_omat} illustrates the electrostatic potential drop close to the interface for both Cheatham's parameters (with standard mixing rules and when optimized for gold-ion interactions) and for the UMA-S(OMat) potential. We can see how the reparametrized force field is in better agreement with the potential profile predicted by the MLIP, particularly for negative potentials, where cations dominate the interfacial electrolyte population. At positive potentials, however, the UMA-S(OMat) potential profile shows a sharper drop than Cheatham, crossing 0~V at a distance of 0.5~\AA{} from the gold. The negative local electrostatic potential -- even if the total potential drop between the interface and the bulk is +0.5~V -- leads to the presence of a charge-compensating diffuse layer of \ce{Na+} ions. This feature, which explains the high differential capacitance observed in the UMA-S(OMat) \ce{NaCl}, is completely absent when using Cheatham's free energy profile with standard mixing rules and only partially present in the reparametrized case.

In this section we discussed how different force field interactions can propagate into markedly different predictions for macroscopic observables such as the potential of zero charge and differential capacitance. This serves as an additional argument supporting the need for an accurate description of interfacial energetics and dynamics. Our results emphasize how reparameterization of classical force fields, guided by higher-level theory, can be an effective method to improve the classical modeling of electrochemical interfaces. The UMA-S(OMat) potential was used here as a surrogate model for higher-level theory, and its ability to properly describe interfacial systems (which is outside its training scope) still needs to be proven. However, the stability of the MLIP to perform enhanced sampling simulations over a long timescale and its ability to predict experimentally sound results are encouraging, especially in light of possible failures related to poor extrapolation exhibited by other models~\cite{Raja2025}. Future development of MLIPs is a promising path forward for predictive and chemically accurate simulations of complex interfacial phenomena. In this context, linking simulation results with mean-field models can effectively bridge atomic-scale quantities to experimental observables. With our continuum model, we show that single-ion adsorption free energy is an important parameter that modulates the structure of the electric double layer.
\section{\label{sec:conclusions}Conclusions}

By combining molecular dynamics simulations, free energy sampling, and continuum modeling of electrochemical interfaces [aqueous \ce{NaF} and \ce{NaCl} at the \ce{Au}(111) surface], we demonstrate how standard mixing rules for LJ parameters -- $\varepsilon$ and $\sigma$ -- can lead to dramatically different outcomes in terms of ion concentration profiles, potentials of zero charge, and differential capacitance curves. This is despite different LJ parameters taken from previous studies leading to similar bulk electrolyte thermodynamics. To retain the computational efficiency of such simplistic classical models, we use the UMA-S(OMat) MLIP as a surrogate \emph{ab initio} benchmark and propose new LJ parameters for the ion-Au interactions, for each specific starting set of ion LJ parameters, that approximate the adsorption free energy profiles generated using the more accurate MLIP. These ion-Au interaction parameters can be viewed as approximately intrinsic, being relatively insensitive to the starting ion interaction parameters. 

This finding indicates that LJ mixing rules may not be advisable for studies of interfaces. However, LJ potentials with corrected interfacial interactions may still provide useful predictions of electrochemical observables, provided that all local minima in the ion adsorption free energy profile can be reproduced. This is shown to be the case for \ce{NaF} solutions. However, systematic differences in differential capacitance behavior at large positive potentials are observed for \ce{NaCl} solutions, primarily due to missing specific adsorption of \ce{Cl-} in the classical LJ models. Future studies may consider correcting for these shortcomings by also updating other force field parameters (ion-water or water-electrode) or by introducing different analytic functional forms for the ion-electrode interaction. Nevertheless, this work establishes a robust protocol for future work on integrating molecular insights into continuum models of the electric double layer through atomistic free energy sampling.

\red{The general applicability of the fitting method described in this work is resilient towards changes of the surrogate model we are trying to match. In this regard, because of the fast-paced development of MLIP, we expect UMA-S(OMat) to be superseded by more accurate models. For example, at the time of writing the conclusions of this manuscript, the newer model version of UMA (1.1) had already been published. At the same time, the inability of UMA-S(OMat) to account for explicit atomic charges and the relatively small cutoff radius for the pairwise interaction (6~\AA{}), make it more a proof of concept that such a generic MLIP can capture qualitative trends in interfacial ion adsorption rather than a quantitatively predictive tool for electrochemical systems where long-range electrostatics and explicit charge transfer play a critical role. Yet, the superior computational performance of MLIPs with respect to AIMD makes them ideal for tasks that require longer trajectories and extensive sampling of the free energy space. The use of MLIPs trained on high-level \textit{ab initio} data, including explicit treatment of electronic polarization and charge transfer, represents a promising avenue to overcome the limitations of classical potentials.}

\FloatBarrier

\section*{Supplementary Material}
The Supplementary Material includes benchmarking data for the MLIPs, an assessment of mixing-rule effects on the adsorption free energy in classical MD, and the spatial profile of the z-dependent relative permittivity of water used in the continuum model. The code used to generate the initial interfacial structures is available at \url{https://github.com/roncofaber/mdinterface}.

\section*{Author Contributions}
All authors helped conceptualize the original idea from FR and AF. FR ran the MD calculations, analyzed the data, and wrote the original manuscript draft with support from AF. AF helped benchmark the MLIP potentials. DP and FR developed and wrote the continuum model framework. All authors discussed the results and contributed to the final version of the manuscript. DP and AM supervised and managed the project. TY secured the funding.

\section*{Conflicts of interest}
There are no conflicts to declare.

\section*{Acknowledgements}
This material is based upon work supported by the U.S. Department of Energy, Office of Science Energy Earthshot Initiative as part of the Center for Ionomer-Based Water Electrolysis at Lawrence Berkeley National Laboratory under Contract DE-AC02-05CH11231 and Basic Energy Science under Award number DE-SC0024239. This research used resources supported by a User Project at The Molecular Foundry and its computing resources, managed by the High-Performance Computing Services Group at Lawrence Berkeley National Laboratory (LBNL), supported by the Director, Office of Science, Office of Basic Energy Sciences, of the United States Department of Energy under Contract DE-AC02-05CH11231. This research used resources of the National Energy Research Scientific Computing Center (NERSC), a Department of Energy User Facility, using NERSC award BES-ERCAP 0033472.

% NERSC + DOE FAIR

%%%END OF MAIN TEXT%%%

%The \balance command can be used to balance the columns on the final page if desired. It should be placed anywhere within the first column of the last page.

\balance

\bibliography{bibliography}% Produces the bibliography via BibTeX.
\bibliographystyle{bibstyle}

%%% ADD SUPP MAT
%\FloatBarrier
%\input{SuppMat/suppmat}

\end{document}